\pdfoutput=1
\documentclass[a4paper,fleqn,usenatbib]{mnras}

\usepackage[T1]{fontenc}
\usepackage{ae,aecompl}
\usepackage{subfig}

\usepackage{graphicx}	% Including figure files
\usepackage{amsmath}	% Advanced maths commands
\usepackage{amssymb}	% Extra maths symbols

\graphicspath{{Plots/}}
\title[Phase-space structure of tidally stripped halos]{The phase-space structure of tidally stripped halos}

\author[N. E. Drakos et al.]{
	Nicole E. Drakos$^{1} $\thanks{E-mail: ndrakos@uwaterloo.ca},
	James E. Taylor$^{1}$\thanks{E-mail: taylor@uwaterloo.ca},
	and Andrew J. Benson$^{2}$
	\\
	$^{1}$Department of Physics and Astronomy, University of Waterloo, 200 University Avenue West, Waterloo, ON, N2L\,3G1, Canada \\
	$^{2}$Carnegie Observatories, 813 Santa Barbara Street, Pasadena, CA 91101, USA\\
}

\date{Accepted XXX. Received YYY; in original form ZZZ}

\pubyear{2016}

\begin{document}
	\label{firstpage}
	\pagerange{\pageref{firstpage}--\pageref{lastpage}}
	\maketitle
	
	% Abstract of the paper
	\begin{abstract}
We propose a new method for generating equilibrium models of spherical systems of collisionless particles that are finite in extent, but whose central regions resemble dark matter halos from cosmological simulations. This method involves iteratively removing unbound particles from a Navarro-Frenk-White profile {\color{black} truncated} sharply at some radius. The resulting models are extremely stable, and thus provide a good starting point for $N$-body simulations of isolated halos. We provide a code to generate such models for NFW and a variety of other common density profiles. {\color{black} We then develop an analytic approximation to this truncated distribution function. Our} method proceeds by analogy with the King model, truncating {\color{black} and shifting} the original distribution function of {\color{black} an infinitely extended} Navarro-Frenk-White profile in energy space. We show that the {\color{black} density profiles of our} models closely resemble the tidally truncated density profiles seen previously in studies of satellite evolution. Pursuing this analogy further with a series of simulations of tidal mass loss, we find that our models provide a good approximation to the full distribution function of tidally stripped systems, thus allowing theoretically motivated phase-space calculations for such systems. 
	\end{abstract}
	
	% Select between one and six entries from the list of approved keywords.
	% Don't make up new ones.
	
	%%%%%%%%%%%%%%%%%%%%%%%%%%%%%%%%%%%%%%%%%%%%%%%%%%
	
	%%%%%%%%%%%%%%%%% BODY OF PAPER %%%%%%%%%%%%%%%%%%
	
\section{Introduction} \label{sec:Intro}

On large scales, a number of independent and complementary tests, including the spectrum of fluctuations in the microwave background \citep[e.g.][]{planck2016}, galaxy clustering \cite[e.g.][]{ BOSS2016}, and weak gravitational lensing \cite[e.g.][]{kitching2014}, provide overwhelming support for the existence of dark matter. This component dominates over regular matter by a factor of $\sim 5\ $ \citep{planck2016}, and collapsed, virialized dark matter halos are thought to be the site of all galaxy formation \cite[see][for a review]{frenk2012}.

On smaller scales, observational tests including galaxy kinematics (e.g.~Ouellette et al. in prep.,\ \citealt{battaglia2013}), satellite kinematics \cite[e.g.][]{prada2003,guo2012}, and weak or strong gravitational lensing \cite[e.g.][]{okabe2013,umetsu2016}, amongst others, are beginning to probe the structure of individual dark matter halos, placing direct constraints on their density profile and velocity structure, as well as central concentration, shape, and substructure. For the moment, however, most of our knowledge of these non-linear scales comes from numerical simulations, which have studied the formation and evolution of halos at ever increasing resolution \citep[e.g.][]{diemand2007, springel2008, diemand2008, stadel2009, gao2012}.

One of the main results of these simulations has been the observation that halos have a universal density profile (UDP), when averaged spherically. The classic approximation to this profile is from the work of \citet[][-- NFW hereafter]{navarro1996,navarro1997}:
\begin{equation}
\rho(r) = \dfrac{\rho_0}{r/r_s(1 + r/r_s)^2} \,\,\, ,
\end{equation}\label{eq:NFW}
where $\rho_0$ is a characteristic density and $r_s$ is the scale radius, corresponding to the point where the logarithmic slope is $d\ln\rho/d\ln r = -2$. This profile has no outer limit {\it a priori}; as $r$ goes to infinity the central potential remains finite but the mass diverges. In a cosmological context, halos are usually considered out to the virial radius, the radius within which they are in approximate virial equilibrium. In practice, the virial radius can be defined by one of several overdensity criteria, or it can be specified in terms of the concentration parameter $c \equiv r_{vir}/r_s$. 

More recent work at higher resolution has determined that halos differ slightly but systematically from the original NFW fit, and that a better approximation is the Einasto profile \citep{navarro2004, merritt2006,gao2008}:
\begin{equation}
\rho(r) = \rho_{-2} \exp \left(-\dfrac{2}{\alpha}\left[\left(\dfrac{r}{r_{-2}}\right)^\alpha - 1 \right]\right) \,\,\, , 
\end{equation}
which contains an extra shape parameter $\alpha$ that seems to vary systematically with mass and redshift \citep{gao2008,dutton2014,klypin2016}. In this paper we will consider only the simpler NFW model, although our results are easily extended to an Einasto profile, and our initial condition code, described in the Appendix, includes the Einasto profile as an option.

Since the discovery of the universal density profile, there has been extensive work characterizing halo properties such as shape,  concentration, spin and substructure (see \citealt{taylor2010} for older references, or \citealt{klypin2016} for more recent references). Average values for properties such as concentration are now well determined as a function of mass, redshift and cosmology. On the other hand, given the complexity of halo growth through hierarchical merging in a cosmological context, idealized, isolated simulations of mergers between pairs of halos may be better suited to determining the physical mechanisms by which these average trends are established. For this work, it is convenient to be able to construct models whose density {\color{black} is truncated} at a finite radius {\color{black} and/or whose mass converges to a finite value}. Prior work \citep[e.g.][]{moore2004,kazantzidis2004} has used models with exponentially truncated NFW profiles, but there is no particular theoretical motivation for this choice. Our initial goal at the outset of this work was to provide a better motivated model for the initial conditions for simulations of isolated, truncated systems.

The standard theory of cosmological structure formation also predicts that halos should grow through repeated, hierarchical mergers, and that the cores of smaller merging systems should survive as tidally-stripped, self-bound substructure within galaxy, group and cluster halos. 
In previous work, \citet{hayashi2003} studied the evolution of halo substructure by simulating the tidal stripping of a smaller satellite halo by a larger host halo. They found that, independent of orbit, the density profile of the satellite halo changed in a predictable way. This change could be described by an empirical model which only required one additional parameter, {\color{black} equivalent to} a tidal radius $r_t$. Though this paper did a thorough job of describing the profile of tidally stripped halos, {\color{black} it did not} determine the exact mechanisms behind the changes. 
Subsequent work by \citet{Kampakoglou2007} showed that tidal forces acting on individual particles may explain some of the form of the {\color{black} density profile of stripped systems, while \cite{choi2009} showed that particles become unbound based more on their energy than on their angular momentum. Despite this work, the net effect of tidal forces on the full distribution function remains somewhat unclear.} 

Observational tests of these theoretical predictions based on lensing \citep[e.g.][]{grillo2015,jauzac2016} or internal kinematics (e.g.~Ouellette et al.~in prep.) are still in their infancy, but show promise given forthcoming datasets from very large surveys. Other substructure calculations, such as the boost factor for dark matter annihilation \citep[e.g.][and references therein]{anderhalden2013,sanchez2014}, depend not only on the density profile, but also on the full distribution function (DF) of dark matter in stripped systems. For these applications, and in order to understand the mechanisms of tidal stripping fully, it would be useful to have a simple analytic model for the DF of a tidally stripped system.

In our experiments with {\color{black} the truncation of distribution functions (DFs)} in energy space,
%, inspired by the truncated star cluster models of King and others, 
we have found a simple method for realizing spatially truncated halos whose central regions resemble NFW profiles. In this paper we describe the method, but also show that the truncated systems it produces closely resemble tidally stripped subhalos orbiting within a larger potential. Thus our model can be used to represent either stable, isolated systems with a finite extent (e.g.~ for merger simulations) or tidally stripped systems (e.g.~for substructure calculations).

The outline of the paper is as follows: in Section 2, we describe an iterative algorithm for generating the ICs for an isolated, {\color{black} spatially truncated} NFW profile. In Section~\ref{sec:model}, we compare this approach to an analytic model where the NFW DF is truncated in energy and shifted, analogously to a King model. In Section~\ref{sec:sat}  we describe our simulations of tidal stripping of halos on various orbits, and in Section~\ref{sec:result} we compare the properties of the stripped remnants to the analytic truncated models. Finally, in Section~\ref{sec:conc} we discuss the implications of our results and summarize our main conclusions.

%%%%%%%%%%%%%%%%%%%%%%%%%%%%%%%%%%%%%%%%%%%%%%%%%%
%%%%%%%%%%%%%%%%%%%%%%%%%%%%%%%%%%%%%%%%%%%%%%%%%%
%%%%%%%%%%%%%%%%%%%%%%%%%%%%%%%%%%%%%%%%%%%%%%%%%%
	
\section{Creating Truncated Initial Conditions} \label{sec:ICs}

To simulate how a spherical dark matter halo evolves in isolation,  we require a method for generating initial conditions (ICs) for particles which collectively form a stable self-gravitating system. Given an analytic expression for the density profile, we can integrate this to obtain the normalized cumulative mass distribution $M(<r)/M_{tot}$, and select particles randomly from this distribution, mapping enclosed mass fraction to radius. Clearly, profiles whose total mass $M_{tot}$ diverges at large radii need to be truncated in some way. The most common solution for cosmological halos \citep[e.g.][]{kazantzidis2004, moore2004, kazantzidis2006, penarrubia2010} is to use the exponentially truncated NFW profile first introduced by \cite{springel1999}, although this form is motivated more by mathematical convenience than by {\color{black} any} physical {\color{black} argument}.

There are two approaches to determining particle velocities (assumed in the simplest case to be spherically symmetric and isotropic in velocity space). We can either calculate the velocity dispersion at each radius from the Jeans equation, making the approximation that the velocity distribution is Maxwellian, \citep[e.g.][]{hernquist1993}, or, for a more accurate model, we can draw particle energies from the full DF \citep[e.g.][]{kazantzidis2004} and convert these to velocities.  We review the latter approach below, before introducing and testing our truncation method.

\subsection{Generating Initial Conditions from the Distribution Function} \label{sec:DF}

We will briefly review the main properties of the DF; a more detailed explanation can be found in \cite{binney}. The DF, $f$, describes the mass per phase-space volume.  {\color{black} For a spherical, isotropic, self-gravitating system, it can be written as a function of a single variable, say a relative energy $\mathcal{E}$.} Given a potential $\Phi$, we will define the {\color{black} (potential and total)} relative energies as $\Psi = -\Phi +\Phi_0$ and $\mathcal{E} = \Psi - v^2/2$. The parameter $\Phi_0$ is free, and is generally set to the value of the potential at the outer boundary of the system, such that the $f>0$ only when $\mathcal{E}>0$, while $f=0$ when $\mathcal{E} \leq 0$.

A relationship between the density profile of a system and the DF can be found by integrating $f(\mathcal{E})$ over all velocities. For a spherically symmetric system:
\begin{equation} \label{eq:recovdens}
\rho(r)=4 \pi \int_0^{\Psi(r)} f
(\mathcal{E}) \sqrt{2(\Psi(r)-\mathcal{E})}d\mathcal{E} \,\,\, .
\end{equation}
Equation \eqref{eq:recovdens} can be inverted \citep{Eddington1916}, and thus the DF can be expressed in terms of the density:
\begin{equation} \label{eq:eddingtoneq}
f(\mathcal{E})=\dfrac{1}{\sqrt{8}\pi^2}\left[ \int_0^\mathcal{E} \dfrac{1}{\sqrt{\mathcal{E}- \Psi}}\dfrac{d^2 \rho}{d \Psi^2} d\Psi +\dfrac{1}{\sqrt{\mathcal{E}}}\left( \dfrac{d\rho}{d\Psi}\right)_{\Psi=0}  \right] \,\,\, .
\end{equation}
A particle at radius $r$ then has energy $\mathcal{E} \in (0,\Psi(r))$ with a probability proportional to $f(\mathcal{E}) \sqrt{\Psi(r) - \mathcal{E}} $.

{\color{black} Given a density profile $\rho$, the relative potential $\Psi(r)$ can be calculated using Poisson's equation, and the derivatives $d\rho/d\Psi$ and $d^2\rho/d\Psi^2$ can be evaluated analytically or numerically. The distribution function $f(\mathcal{E})$ can than be calculated for a given relative energy $\mathcal{E}$ using Equation (\ref{eq:eddingtoneq}). To generate a model halo, a radius and a relative energy are selected at random for each particle, in such a way as to reproduce the correct density profile and distribution function.} Once {\color{black} the radius and relative energy} have been assigned, the norm of the velocity of each particle can be calculated as {\color{black} $v = \sqrt{2(\Psi(r) - \mathcal{E})}$}. Finally, 3D position and velocity {\color{black} components} can be chosen at random assuming (spatial) spherical symmetry, and isotropy in velocity space, respectively. {\color{black} Although these techniques are well known, for convenience we review them in Appendix A.}

\subsection{Truncating the NFW Profile at a Finite Radius} \label{sec:IC_method}

The exponentially truncated NFW profile was first described in \cite{springel1999}, and is identical to the NFW profile within the virial radius, $r_{vir}
$, but is truncated exponentially beyond that. How fast this decay occurs depends on the parameter $r_{d}$:
\begin{equation}
\rho(r) = \begin{cases} \dfrac{\rho_0 r_s^3}{r(r_s+r)^2} &\mbox{if } r<r_{vir} \\
\dfrac{\rho_0}{c(1+c)^2} \left( \dfrac{r}{r_{vir}} \right)^{\epsilon} \exp \left(- \dfrac{r-r_{vir}}{r_d}\right)& \mbox{if } r>r_{vir} \end{cases} \,\,\, , 
\end{equation}
where the constants $\rho_0$ and $r_s$ are as in Equation \eqref{eq:NFW}.  Additionally, to ensure that the logarithmic slope at $r_{vir}$ is continuous, there is the constraint:
\begin{equation}
\epsilon = - \dfrac{r_s+3r_{vir}}{r_s+r_{vir}} + \dfrac{r_{vir}}{r_d} \,\,\, .
\end{equation}
Not only does this solution have little physical motivation, {\color{black} however, but} it is also discontinuous in the second derivative of {\color{black} $\rho(r)$}. Since Equation \eqref{eq:eddingtoneq} depends on this second derivative through the ${d^2 \rho}/{d \Psi^2}$ term, the DF may not be monotonically increasing for certain choices of $r_d$, resulting in unphysical behaviour.

{\color{black} Instead}, we propose a different modification to the NFW profile. Our solution is to consider particles as if they were sampled from an infinitely extended NFW profile, but only choose those within some radius $r_{cut}$. Particle energies are initially assigned from the {\color{black} DF corresponding to the} full (infinitely extended) NFW {\color{black} profile}, as described in Section~\ref{sec:DF}. We then iteratively remove any unbound particles, using at each step of the iteration the potential defined by the set of particles remaining.\footnote{\color{black}{Note \cite{choi2007} propose an alternative, but more complicated method, truncating the initial density profile, calculating a distribution function via Eddington inversion, recalculating the resulting density profile, and iterating over these steps until convergence.}}  For a halo with $r_{cut} = 10\,r_s$, this process converges within 10 iterations or fewer, leaving approximately 65\% of the mass within $r_{cut}$ bound. The final result is a system with a density profile that matches NFW at small radii, but drops off more steeply at large radii, reaching zero density at $r_{cut}$. While this truncated profile is instantaneously self-bound, its long term stability is unclear. We will consider this point in the next section.

\subsection{Stability Tests}

Before we test the stability of our truncated profile, we need to estimate the possible contribution from collisional effects. These can complicate the interpretation of any N-body simulation. Our method for determining particle ICs assumes that the particles exist in a smooth continuous potential; each particle can be considered a Monte Carlo sampling of this potential. However, individual particle-particle interactions lead to a gradual loss of the initial structure, most noticeably in the dense center of the halo. 

There are two time scales on which collisional effects are important. The first, the relaxation time scale, $t_{rel}$, corresponds to the time which it takes a typical particle's velocity to change by an order of itself. The second, slower timescale is the evaporation time, $t_{evap}$, the amount of time it takes a typical particle to reach escape speed, and thus `evaporate' from the system. Following the arguments from \cite{binney}, we calculate:
\begin{equation}
\begin{aligned}
t_{rel} (r) &\approx 0.1 \dfrac{\sqrt{N(<r)}}{\ln N(<r)}  \sqrt{\dfrac{r^3}{Gm}} \,\,\, , \mbox{     and}\\
t_{evap} (r)&\approx 136\, t_{rel} (r)  \,\,\,,
\end{aligned}
\end{equation}
where $N(<r)$ is the number of particles within radius $r$. Using these formulae, we can define a `(central) relaxation radius', $r_{relax}(t)$, and a `(central) evaporation radius', $r_{evap}(t)$, such that $t_{rel}(r_{rel}) = t$ and $t_{evap}(r_{evap}) = t$ respectively, within which relaxation and evaporation are important.

To verify the stability of our initial satellite halo, we evolve it in isolation with the $N$-body  code \textsc{Gadget-2} \citep{gadget2}. We generate an initial NFW system with  {\color{black}$2 \times 10^6$} particles within a radius $r_{cut}= 10 r_s$. After iteratively removing unbound particles (as described in Section~\ref{sec:IC_method}), the final number of particles in the truncated system is {\color{black}$N=1286991$}. We then assign each particle a mass $m$ such that the total satellite has mass $M_{sat}$. We simulate the evolution of the system using the softening length proposed in \cite{van2000}, $\epsilon = 0.5 r_h N^{-1/3} $, where $r_h$ is the half-mass radius. For our ICs,  {\color{black} $\epsilon = 0.01 r_s$}.

\begin{figure}
	\includegraphics[width=\columnwidth]{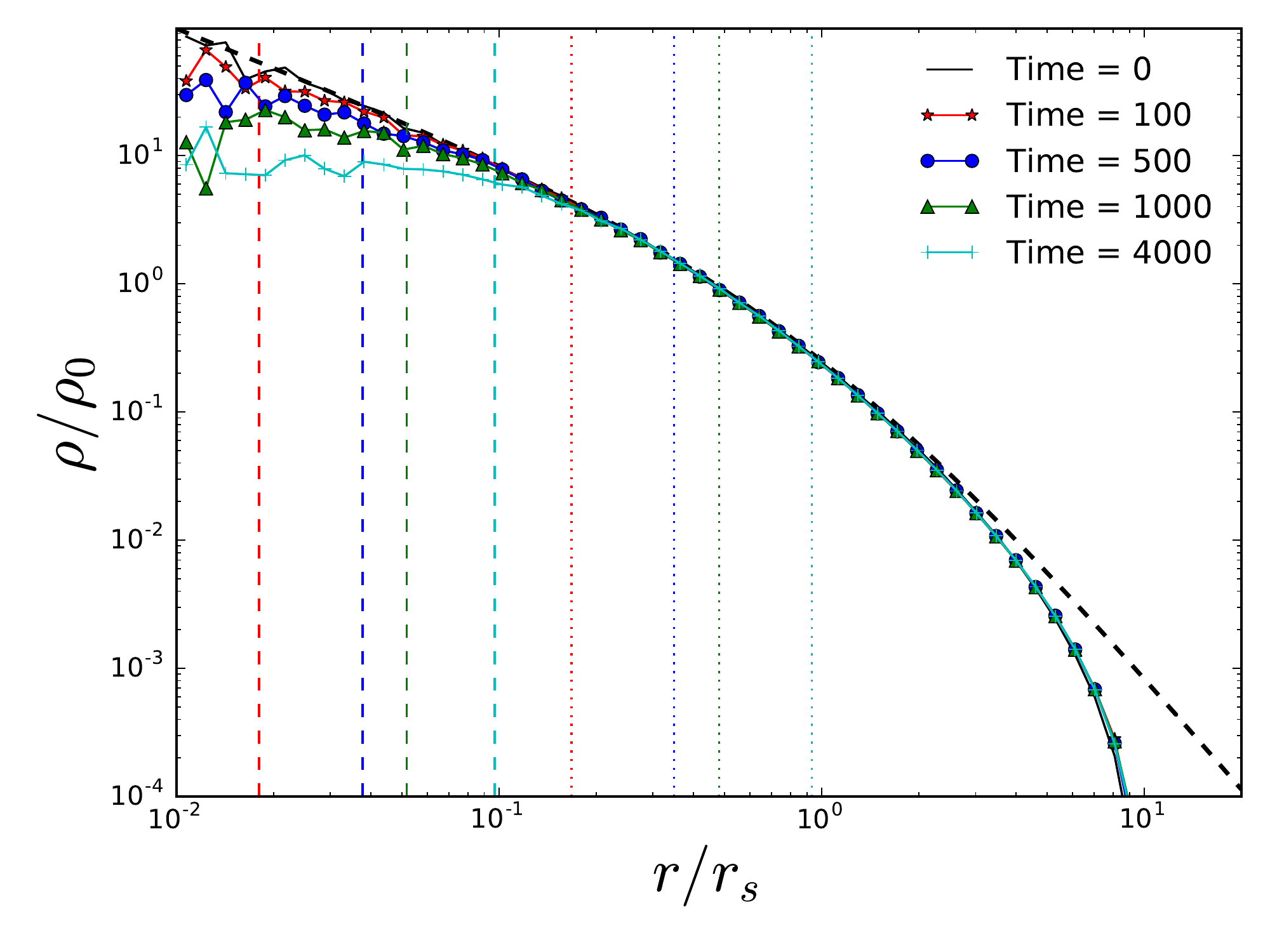}
	\caption{Evolution of the density profile simulated in isolation. The black dashed line shows a (non-truncated) NFW profile, and the {\color{black} solid black} line shows the profile of our ICs at $t=0$. The coloured lines show the density profile at subsequent times, as labeled. Radii $r_{rel}$ and $r_{evap}$ at which relaxation and evaporation effects become important are shown with vertical dotted and dashed lines, respectively. The truncated profile is completely stable outside $r_{evap}$. Time is in units of $\sqrt{r_s^3/GM_{sat}}$, where $r_s$ is the scale radius, and $M_{sat}$ is the mass of the halo.}
	\label{fig:IC_profile}
\end{figure}

Fig.~\ref{fig:IC_profile} shows the evolution of the density profile with time. The outer part of the profile appears to be completely stable, with any systematic changes invisible on the scale of the figure. The only visible change is in the region interior to $r_{evap}$. {\color{black}This is} consistent with the results {\color{black} of} \cite{hayashi2003}, who found {\color{black} that} the internal structure evolves at radii where the evaporation time is close to $t$. {\color{black} (We have also confirmed the predicted scaling of $r_{evap}$ with particle number using lower resolution simulations.)} We conclude that this central change is a collisional effect due to finite resolution, and that our ICs are otherwise extremely stable.

One disadvantage of our iterative unbinding method is that the resulting truncated DF does not have a simple expression, even for cases where the original, un-truncated DF is simple. Thus, in the next section we will consider a slightly different analytic approach, and show that it produces remarkably similar results.

%%%%%%%%%%%%%%%%%%%%%%%%%%%%%%%%%%%%%%%%%%%%%%%%%%
%%%%%%%%%%%%%%%%%%%%%%%%%%%%%%%%%%%%%%%%%%%%%%%%%%
%%%%%%%%%%%%%%%%%%%%%%%%%%%%%%%%%%%%%%%%%%%%%%%%%%
\section{An Analytic Model for the Truncated Distribution Function} \label{sec:model}

Our goal is to derive an analytic approximation for the DF of a truncated system with a NFW or UDP-like central density profile. We will take an approach similar to the derivation of the King model \citep{king1966}, {\color{black} truncating and lowering the DF in energy space, as} described below. {\color{black} For the NFW profile, this approach has been proposed previously by \cite{widrow2005}, who showed that it leads to sharply truncated density profiles (see their Figure 1).}

\subsection{Review of the King Model} \label{sec:kingsec}

The King model is a lowered isothermal model based on the (infinitely extended) isothermal sphere, but with a lowered relative energy $\mathcal{E}$. Additionally, it subtracts a constant term to ensure that the DF is continuous at this truncation energy. This last step results in a more extended profile \citep{hunter1977}, and presumably increases the stability of the model \citep[e.g.][]{Guo2008}.

As previously, we will define relative energies $\Psi = -\Phi +\Phi_0$ and $\mathcal{E} = \Psi - v^2/2$. Suppose initially we choose $\Phi_0 = 0$. The DF for an isothermal sphere of velocity dispersion $\sigma$ can be written:
\begin{equation} \label{eq:iso}
F_{iso}(Z) = F_0 \exp[Z] \,\,\, ,
\end{equation}
where $Z = \mathcal{E}/\sigma^2$ {\color{black} is a dimensionless, scaled version of the relative energy}.
As with the NFW profile, this DF corresponds to a profile that extends to infinity and has infinite mass.

A solution to the infinite mass problem of the isothermal sphere is to lower the relative energy, letting $Z \rightarrow Z - Z_t$  \citep{Woolley1954}. This produces a truncated DF of the form:
\begin{equation} \label{eq:wooley}
F_{Wooley}(Z) = 
\begin{cases}
F_{iso}(Z - Z_t) & Z \ge Z_t \\
0 & Z\le Z_t \,\,\,.
\end{cases}
\end{equation}
If {\color{black} one uses} the freedom of $\Phi_0$ to express {\color{black} $F$ in terms of} $Z' =Z - Z_t$, the DF has the same form as {\color{black} that of the infinitely extended} isothermal sphere, but {\color{black} the density now drops to zero at} a finite radius.

This form of the DF introduces a new complication, {\color{black} however,} as it is now discontinuous at $F(Z_t)$. A solution to this {\color{black} problem} is to subtract a constant term. DFs of this form are known as King models \citep{michie1963, king1966}:
\begin{equation} \label{eq:King}
F_{King}(Z) = 
\begin{cases}
F_{iso}(Z - Z_t) - F_{iso}(0) & Z \ge Z_t \\
0 & Z\le Z_t \,\,\,.
\end{cases}
\end{equation}

\subsection{Energy Truncation of a NFW Distribution Function}

For mathematical convenience, we introduce the following dimensionless variables; $R=r/r_s$, $p=\rho/\rho_0$, $P=\Psi/(4\pi G \rho_0 r_s^2)$, $Z=\mathcal{E}/(4\pi G \rho_0 r_s^2)$ and $F=(4\pi G)^{3/2} r_s^3 \rho_0^{1/2}f$, {\color{black} where $\Psi(r) =-\Phi(r) + \Phi_0$ and $\mathcal{E} =\Psi(r)  - v^2/2 $, as above}. Here, the energies have been normalized by the magnitude of the central potential of a NFW profile $|\Phi_{NFW}(r=0)| = 4\pi G \rho_0 r_s^2$.

The DF for the NFW profile can be determined numerically from Equation \eqref{eq:eddingtoneq}. However, in practice, we use the analytic approximation proposed by \cite{widrow2000}:
\begin{multline}
F_{NFW}({\color{black}Z}) =F_0 Z^{3/2}(1- Z)^{-5/2} \left( -\dfrac{\ln Z}{1-Z}\right)^q \\
\times \exp(p_1 Z + p_2 Z^2 + p_3Z^3 + p_4 Z^4 ) \,\,\, ,
\end{multline}
where $q=-2.7419$, $p_1=0.3620$, $p_2=-0.5639$, $p_3=-0.0859$, $p_4=-0.4912$ and $F_0=0.091968 $. { \color{black} We find that this approximation agrees with the numerically calculated DF to within 2\%.}

We wish to truncate the NFW DF at some truncation energy $\mathcal{E}_t$, or in dimensionless form $Z_t=\mathcal{E}_t/(4\pi G \rho_0 r_s^2)$. {\color{black}The modified DF is then given by:
	\begin{equation} 
	F(Z) = 
	\begin{cases}
	F_{NFW}(Z) - F_{NFW}(Z_t) & Z \ge Z_t \\
	0 & Z\le Z_t \,\,\,.
	\end{cases}
	\end{equation}
We can exploit the freedom of $\Phi_0$, picking a new value such that the relative energy is zero on the boundary. Given this new value of $\Phi_0$, and denoting the new relative  energy variable $Z' = Z - Z_t$, the DF then becomes:
	\begin{equation} \label{eq:DF}
	F(Z') = 
	\begin{cases}
	F_{NFW}(Z'+Z_t) - F_{NFW}(Z_t) & Z' \ge 0 \\
	0 & Z'\le 0 \,\,\, ,
	\end{cases}
	\end{equation}
where $Z_t$ the truncation energy defined using the original value of $\Phi_0$.
}

Note that this derivation is slightly different {\color{black} from that of the lowered isothermal sphere model outlined in Section~\ref{sec:kingsec}. Comparing Equations~\eqref{eq:King} and~\eqref{eq:DF}, we find that applying the latter to the isothermal sphere will recover the King model, but multiplied by a constant term $e^{Z_t}$.}

It can be shown that the {\color{black}relative} energy of the energy-truncated NFW profile has a maximum {\color{black} value} $Z = 1 - Z_t$; this corresponds to the central relative potential $P(0)$ of the 
truncated system. The final system {\color{black} has} a finite radius, $r_t$. Increasing the truncation energy will decrease the central potential, the truncation radius,  and the total mass of the system.

We need two more equations to describe this system. The density profile can be recovered from Equation \eqref{eq:recovdens}, and the relative potential can be determined from Poisson's equation:
\begin{equation} 
\dfrac{d^2 \Psi}{dr^2} + \dfrac{2}{r}  \dfrac{d\Psi}{dr}= -4 \pi  G \rho(\Psi) \,\,\,,
\end{equation}
or, re-expressed in the dimensionless parameters:
\begin{equation}\label{eq:poiss}
\dfrac{d^2 P}{dR^2} + \dfrac{2}{R}  \dfrac{dP}{dR}= -p(P) = - 4\pi \int_0^P F(Z) \sqrt{2(P - Z)}dZ \,\,\, .
\end{equation}
The initial conditions are $P(0) = 1 - Z_t$ and $dP(0)/dR = 0$. Equation \eqref{eq:poiss} can be integrated numerically until $P(R_t)=0$.

Putting all this together, the density of the truncated halo at radius ${\color{black} R}$ can be determined using the following steps:
\begin{enumerate}
	\item Specify the dimensionless truncation energy $Z_t \in (0,1)$.
	\item {\color{black}Given the distribution function,} calculate the potential at radius ${\color{black}R}$ by numerically integrating Poisson's equation (Equation \eqref{eq:poiss}) with initial conditions $P(0) = 1 - Z_t$, $dP(0)/dR = 0$.
	\item Find the density by numerically integrating Equation \eqref{eq:recovdens}.
\end{enumerate}
The potential $\Phi (r) = - \Psi(r) + \Phi_0$ can be recovered once the truncation radius has been determined{\color{black}, since} $\Phi_0 = -GM(r_t)/r_t$. 

Finally, we note that although we have discussed the energy-truncation method specifically for an NFW profile, it can be used for any {\color{black} density} profile with a known or calculable DF, including an Einasto profile{\color{black}, and could also be extended to the various theoretically motivated models of the intrinsic halo DF \citep[e.g.][]{Hjorth2010,Pontzen2013,Beraldo2014}. For cored density profiles, energy truncation will generally reduce the central density of the system significantly as the outer radius decreases, as discussed by \cite{widrow2005} and \cite{penarrubia2010}.}

\subsection{Properties of the Truncated Model} \label{sec:Zt}

\begin{figure}
	\includegraphics[width=\columnwidth]{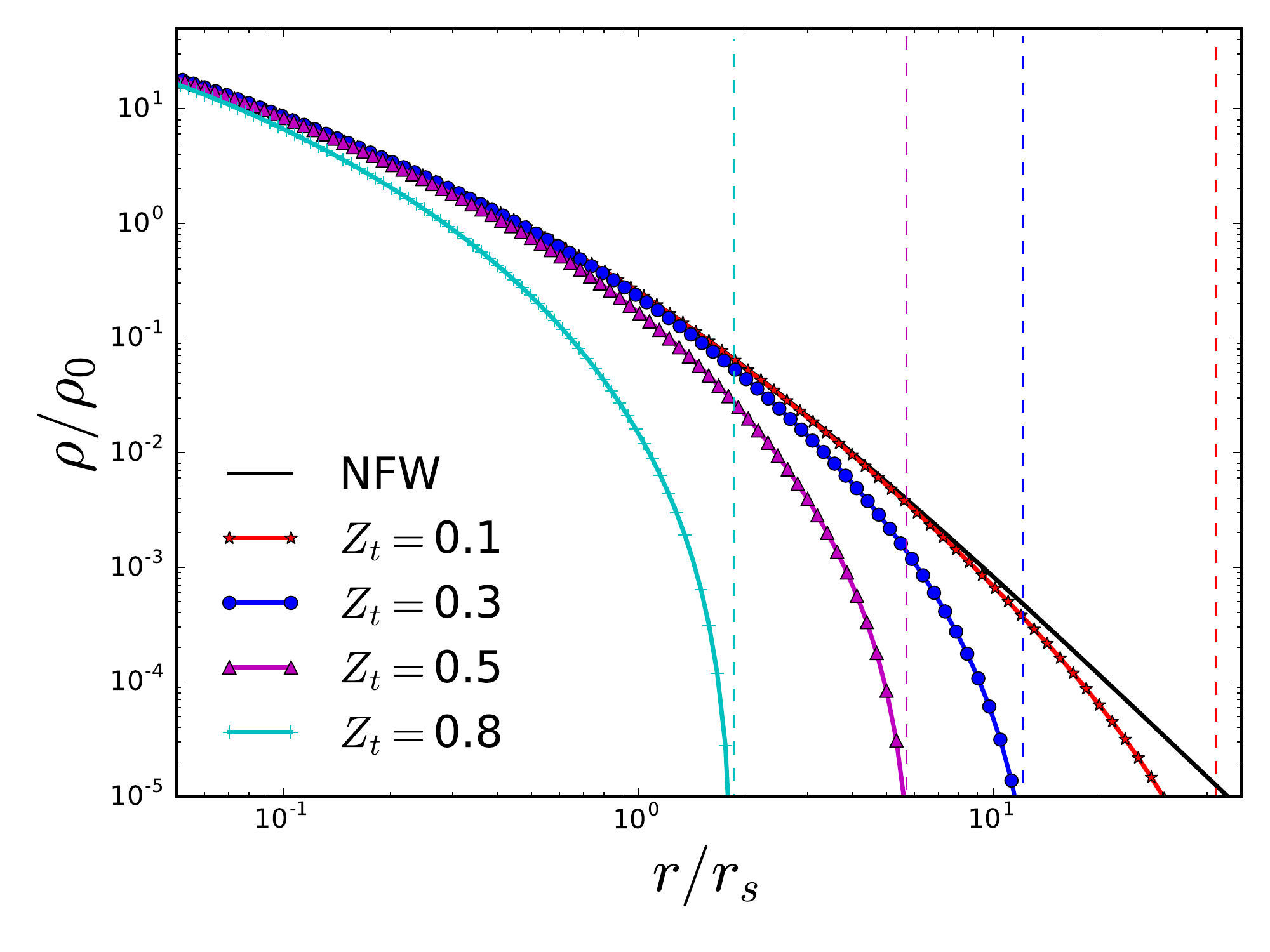}
	\caption{Density profile of an energy-truncated NFW profile. The truncation energy, $Z_t$, is the dimensionless relative energy, $Z_t = \mathcal{E}_t/4 \pi G \rho_0 r_s^2$. Vertical dashed lines indicate the truncation radius, $r_t$.}
	\label{fig:NFWT_dens}
\end{figure}

As we increase the truncation energy, {\color{black} the mass and extent of the NFW profile will decrease progressively}. Fig.~\ref{fig:NFWT_dens} shows how the density profile changes as a function of the dimensionless truncation energy $Z_t$, relative to the original NFW profile. Note that by definition, the density drops to zero outside the truncation radius (indicated by the vertical dashed lines).

\begin{figure}
	\includegraphics[width=\columnwidth]{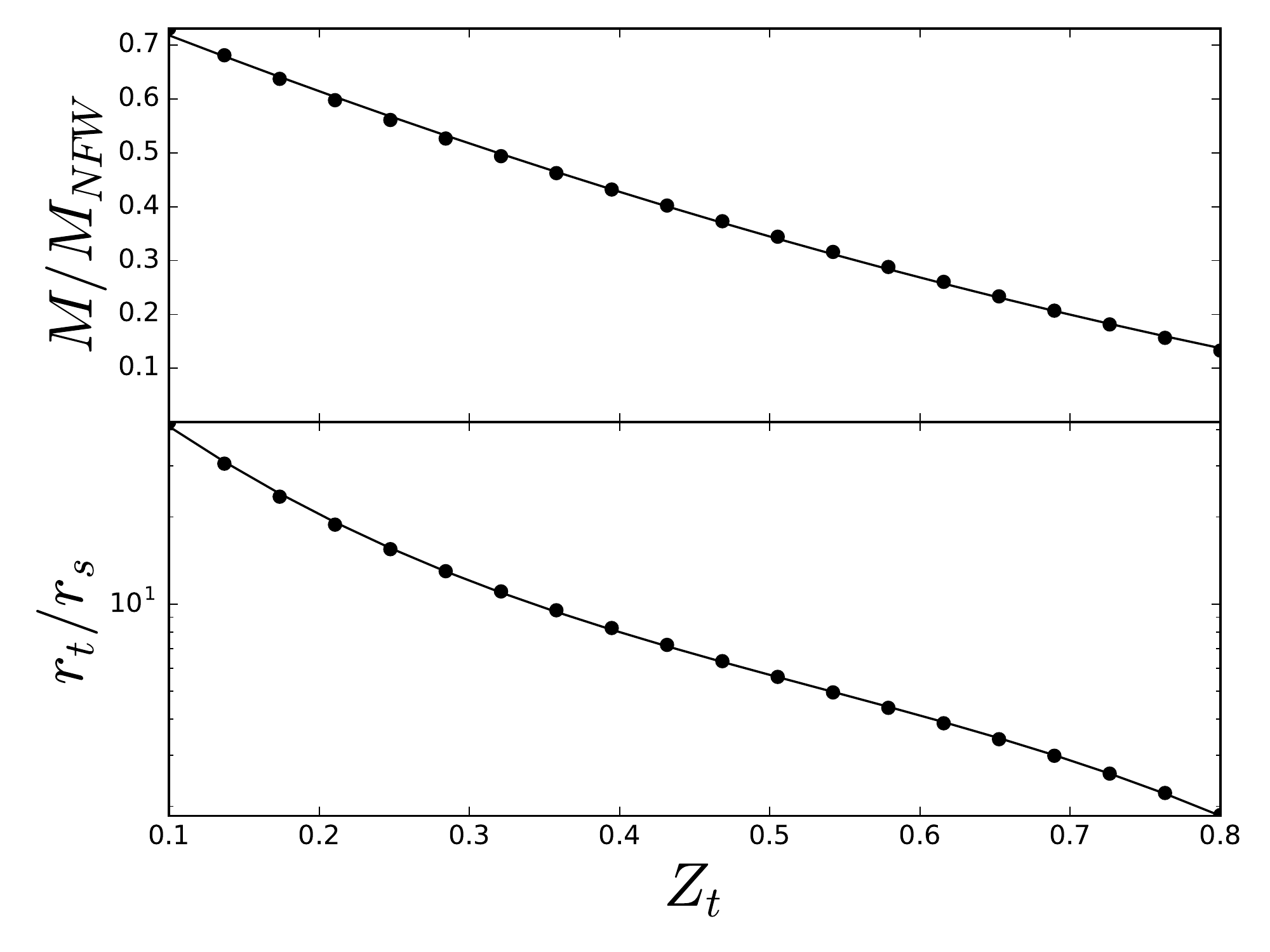}
	\caption{How the bound mass fraction (top) and truncation radius (bottom) change as a function of the dimensionless truncation energy, $Z_t = \mathcal{E}_t/4 \pi G \rho_0 r_s^2$. The bound mass fraction is defined as the total mass within the truncation radius divided by the mass of an untruncated NFW profile within the same radius. Polynomial fits are given in the text.}
	\label{fig:NFWT_Et}
\end{figure}

The relationships between the truncation energy $Z_t$ and the total bound mass and truncation radius are shown in Fig.~\ref{fig:NFWT_Et}. Both the mass and the truncation radius are smooth, decreasing functions of the truncation energy, as expected.  {\color{black} The polynomial fits to the data shown (and valid over the range of the plot) are $M_f \equiv M(r<r_t)/M_{NFW}(r<r_t) =  0.35 Z_t^2 - 1.14 Z_t + 0.83 $ and $r_f \equiv \log_{10}(r_t/r_s)  = -3.70 Z_t^3 + 5.93 Z_t^2 -4.56 Z_t +2.01$. We can also fit the inverse relations: $Z_t = 0.6 M_f^2 -1.7 M_f + 1.02$ and  $Z_t = 0.2 r_f^3 - 0.4 r_f^2 - 0.39 r_f +0.94$.
}

\begin{figure}
	\includegraphics[width=\columnwidth]{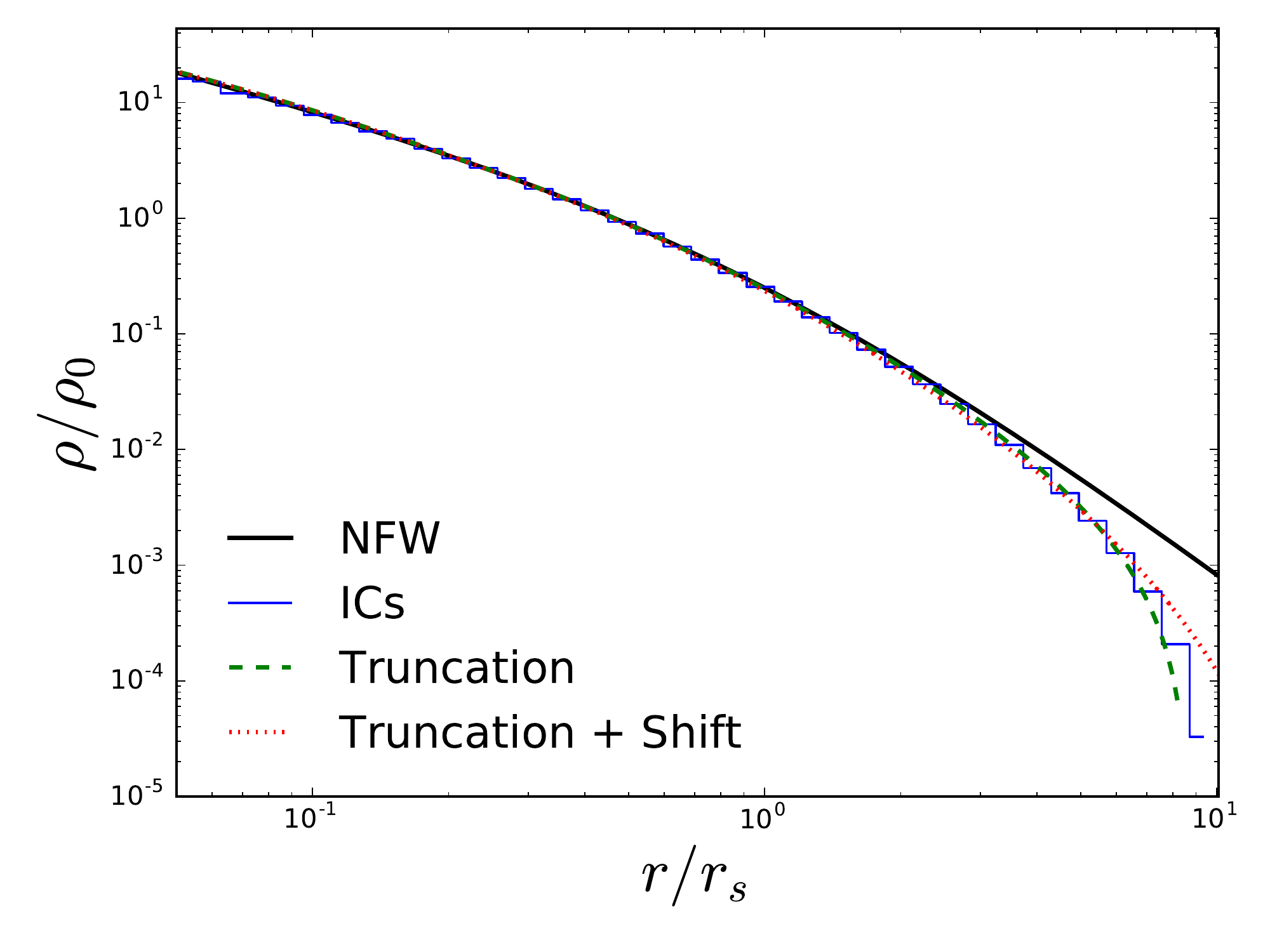}
	\caption{Analytic models for the truncated profile, compared to ICs generated using the iterative method (histogram). The lower {\color{black} dashed} curve shows the result of an energy truncation alone, such that $f(\mathcal{E}) = f_{NFW}(\mathcal{E}+\mathcal{E}_t)$. The upper {\color{black} dotted} curve shows the result of a truncation plus a shift to make the DF continuous at zero: $f(\mathcal{E}) = f_{NFW}(\mathcal{E}+\mathcal{E}_t) - f_{NFW}(\mathcal{E}_t)$. The {\color{black} solid} line shows the untruncated NFW profile.}
	\label{fig:IC_compare}
\end{figure}

Finally, we return to our original goal, to establish {\color{black} an analytic approximation to the DF produced by the iterative unbinding procedure we introduced} in Section~\ref{sec:ICs}. In Fig.~\ref{fig:IC_compare}, we show how the ICs derived in Section~\ref{sec:ICs} compare to the {\color{black} analytic} energy-truncated NFW model developed in this section. The histogram shows the ICs, the lower {\color{black} dashed} curve shows the result of truncating the NFW DF at the energy $\mathcal{E}_t$ (i.e.~$f(\mathcal{E}) = f_{NFW}(\mathcal{E}+\mathcal{E}_t)$), while the upper {\color{black} dotted} curve shows the full model, including the shift to make $f$ continuous at zero: $f(\mathcal{E}) = f_{NFW}(\mathcal{E}+\mathcal{E}_t) - f_{NFW}(\mathcal{E}_t)$. As before, the {\color{black} solid} line shows the original untruncated profile. Both {\color{black} models} provide a good match to the density profile of the ICs. As with the King model, subtracting the constant term $f_{NFW}(\mathcal{E}_t)$ produces a more extended profile. We will adopt this version as our final analytic model for the DF, on the assumption that it is slightly more stable than the model where $f$ is discontinuous at zero.
Overall, these results suggest that an energy-truncated DF comes close to describing {\color{black} ICs obtained using the} method outlined in Section~\ref{sec:ICs}. In principle, ICs could therefore be generated directly from the energy-truncated DF, although in practice our code implements the iterative unbinding procedure.

%%%%%%%%%%%%%%%%%%%%%%%%%%%%%%%%%%%%%%%%%%%%%%%%%%
%%%%%%%%%%%%%%%%%%%%%%%%%%%%%%%%%%%%%%%%%%%%%%%%%%
%%%%%%%%%%%%%%%%%%%%%%%%%%%%%%%%%%%%%%%%%%%%%%%%%%
\section{Simulating Tidally Stripped Halos} \label{sec:sat}

The very similar truncated profiles derived either by iterative unbinding in Section~\ref{sec:ICs}, or analytically in Section~\ref{sec:model}, were designed to represent isolated systems of finite mass and radial extent. We note, however, that they {\color {black} also} look very similar to the {\color {black} density} profiles of tidally stripped halos orbiting within the potential of a larger system \citep[e.g.][]{hayashi2003}. To pursue this analogy, we will compare our truncated models {\color{black} directly} to tidally stripped halos taken from simulations {\color{black} of satellite mass loss}. We describe the simulations and basic analysis {\color{black} below}, and then compare the simulated systems to our models in Section~\ref{sec:result}.

\subsection{Simulation Parameters}

Simulations of a smaller `satellite' halo orbiting within the potential of a larger static `host' halo were performed using the $N$-body code \textsc{Gadget-2} \citep{gadget2}. This code was modified to contain a fixed background potential corresponding to a host halo with an NFW profile. We use the mass and scale radius of the satellite halo ($M_{sat}$ and $r_s$) as the mass and distance units. Time is given in units $t_{unit} =\sqrt{r_s^3/GM_{sat}}$, and velocity in units of $v_{unit} = \sqrt{GM_{sat}/r_s}$. The host and satellite halos were assumed to have the same initial density within their outer, or `virial' radii, as would be the case for a merger between two cosmological halos at a fixed redshift. We generated initial conditions for the satellite using our iterative unbinding algorithm with $r_{cut} = r_{vir} = 10\,r_s$, as described in Section~\ref{sec:ICs}. The virial radius of the main halo scales as $(M_{host}/M_{sat})^{1/3}$, while the scale radius of the host was set assuming $c_{host} = 10$. 

{\color{black}

The orbital parameters of infalling satellite halos have been studied extensively in simulations  \citep{tormen1997a,ghigna1998,vitvitska2002,benson2005,wang2005,zentner2005, khochfar2006,wetzel2011,jiang2015}. 
The parameters used in this paper (see Table~\ref{tab:sims}) cover the full range of energy and angular momentum expected for cosmological mergers \citep[e.g.][]{jiang2015}, with the exception of Simulations 5 and 6  which have unusually low and high energies, respectively. This allows us to test our tidal-stripping model not only for cosmological orbits, but also for a few more extreme cases.

We considered four different host/satellite mass ratios, $M_{host}/M_{sat}=300,\,100,\,50$ and $10$. Since the host halo was modelled as a fixed background potential, the satellite is not subject to dynamical friction in our simulations, and we expect its specific energy and angular momentum to be roughly conserved, even as it loses mass. The assumption that the host halo is static becomes less physically valid for low mass ratios, where dynamical friction plays a larger role. Nonetheless, we run a few cases at smaller mass ratios to test the effect of a larger satellite on the evolution of the density profile. At large mass ratios, where the satellite is small compared to the scale of the background potential, we expect satellite evolution to become independent of mass ratio. We chose a limiting value of $M_{host}/M_{sat}=300$ for practical reasons, as the ratio of the simulation time step to the orbital period is becoming very long at this point.}

\begin{table*}
	\caption{\label{tab:sims} Summary of simulation parameters. Columns give (1) the simulation number (2) the mass ratio between the host and satellite halo (3) the virial radius of the host (4) the apocentric distance (5) the pericentric distance {\color{black} (6) the tangential velocity at apocenter (7) the (radial) orbital period,  (8) the circularity of the orbit, (9) the relative energy (defined as the energy divided by the energy of a circular orbit at the virial radius) (10) the radius of a circular orbit with the same energy divided by the virial radius.}}
	\begin{tabular}{ c c c c c c c c c c}
		\hline
		Simulation & $M_{host}/M_{sat}$& $R_{vir}/r_s$ & $r_a/r_s$ & $r_p/r_s$ &$v_a/v_{unit}$& $t_{orb}/t_{unit}$&  $\epsilon_c$ &  $\eta$ &  $R_c/R_{vir}$ \\ \hline
		1 & 100 & 46.4 & 100 & 10 & 0.34 & 206.8 & 0.42 & 0.85 & 1.26 \\ 
		2 & 100 & 46.4 & 100 & 50 & 0.90 & 299.4 & 0.92 & 0.71 & 1.63\\
		3 & 300 & 66.9 & 100 & 10 & 0.51 & 129.7  & 0.40 & 1.09 &0.88 \\
		4 & 300 & 66.9 & 100 & 50 & 1.42 & 185.4 & 0.92 & 0.92 & 1.13 \\
		5 & 100 & 46.4 & 500 & 50 & 0.23 & 1778.5 &  0.47  & 0.24 & 6.14\\
		6 & 300 & 66.9 & 25 & 10 & 1.50 & 31.48& 0.82  & 2.29 & 0.26 \\
		7 & 50 & 36.8 & 80 & 5 & 0.19 & 201.6& 0.30  & 0.86  & 1.24 \\
		8 & 50 & 36.8& 90 & 15 & 0.37 & 259.7 & 0.58  & 0.76 & 1.48 \\
		9 & 10 & 21.5 & 40 & 10 & 0.30 & 196.8 & 0.71  & 0.88 & 1.19 \\
		10 & 10 & 21.5 & 25 & 10 & 0.42 & 123.2 & 0.85  & 1.14 & 0.82 \\
		\hline
	\end{tabular}	
\end{table*}

\subsection{Locating the Satellite Remnant}

We identified the satellite remnant at any given time using a method similar to the one outlined in \cite{tormen1997,tormen1998}. This is also the method used in \cite{hayashi2003}. There are two steps to this method; first the highest density peak of the particles is located approximately by iteratively decreasing the radius $R$ of a sphere, and re-centering it at each step on the center of mass of the particles contained within the sphere. We decreased the sphere by $0.9R$ on each iteration, and repeated until there were fewer than 100 particles within the sphere. We expect our results to be insensitive to this choice of final particle number, as discussed in \cite{tormen1997}. The velocity of the satellite frame was then calculated as the average velocity of all satellite particles within a sphere of radius $r_{cut}$ {\color{black} (the original truncation radius of the satellite)} centered on the highest density peak. The second step of this process was to identify which subset of particles was self-bound in this frame. This was calculated by iteratively removing unbound particles in the rest frame of the satellite until the algorithm converged. 

Finally, we found that in a few cases where the algorithm had trouble locating a self-bound remnant (particularly at late times in Simulation {\color{black} 6}), we were able to improve the algorithm by first approximating the location of the satellite remnant by integrating {\color{black} forward} from the previous snapshot (assuming the satellite was a point mass orbiting in the potential of the host), and then only considering particles within $2 r_{cut}$ of this predicted location. 

\begin{figure}
	\includegraphics[width=\columnwidth]{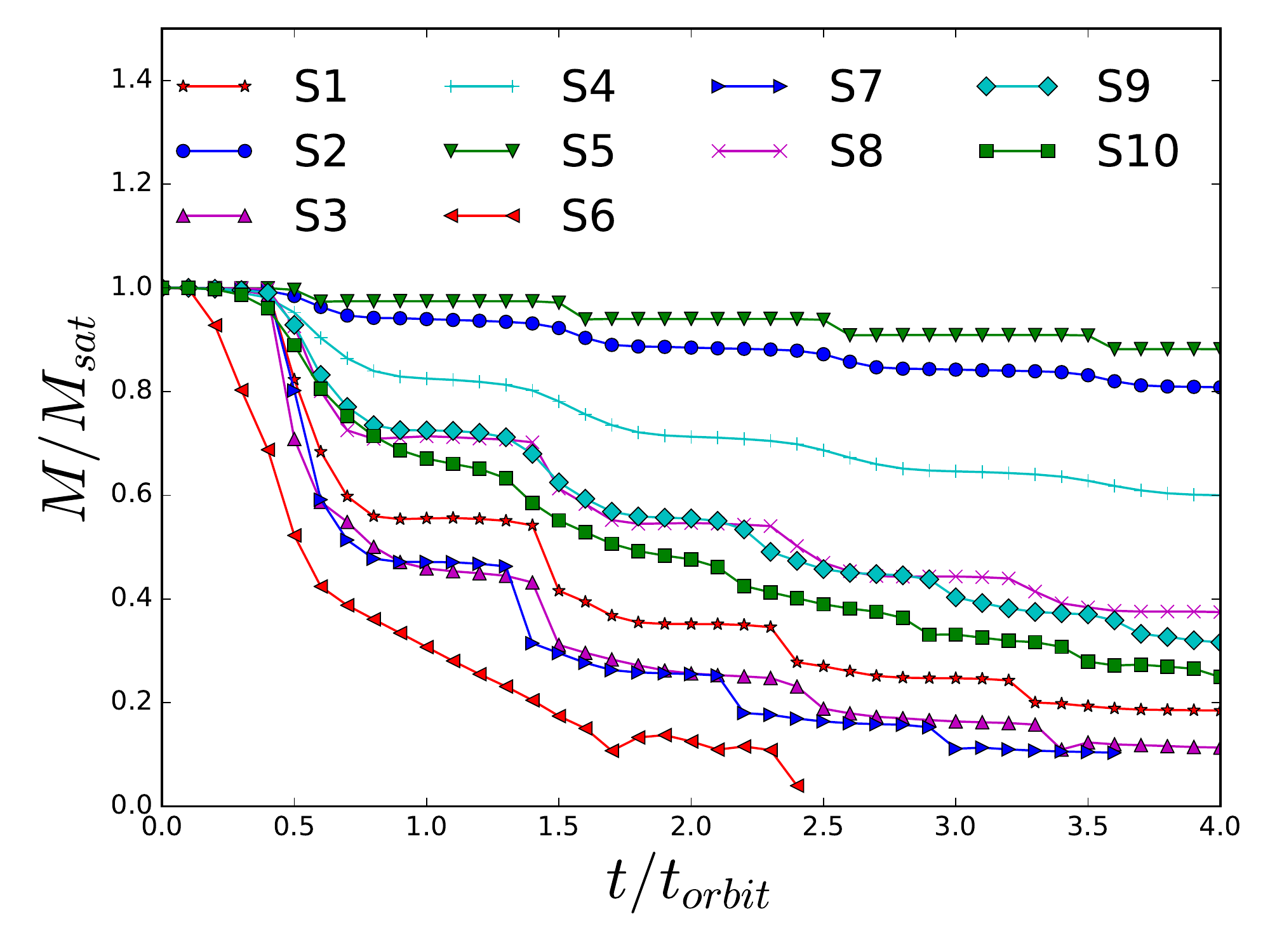}
	\caption{Bound mass fraction as a function of time (in units of the orbital period).}
	\label{fig:time_plots}
\end{figure}

Fig.~\ref{fig:time_plots} shows how the recovered satellite mass decreases with time for the ten simulations.
Given a bound remnant and associated satellite reference frame in each step, we then fit the remnant's profile to the analytic model from Section~\ref{sec:model}, as described below.

%%%%%%%%%%%%%%%%%%%%%%%%%%%%%%%%%%%%%%%%%%%%%%%%%%
%%%%%%%%%%%%%%%%%%%%%%%%%%%%%%%%%%%%%%%%%%%%%%%%%%
%%%%%%%%%%%%%%%%%%%%%%%%%%%%%%%%%%%%%%%%%%%%%%%%%%
\section{Results} \label{sec:result}

\subsection{Fitting Criteria}

To compare our tidally stripped satellites to the energy truncated DF model, we need a way of determining the free parameter, {\color{black} $Z_t$}. We do this by normalizing our DF models, such that their scale radii and central densities before truncation match those of the original NFW profile from which the ICs for the simulations were derived by the iterative method. This accounts for the fact that the initial conditions in the simulations are already truncated at a finite radius. $Z_t$  is then chosen so that both the simulations and the models have the same mass. We compare the analytic model to the simulations at apocenter on successive orbits, since these are the times when we expect the satellite to be closest to equilibrium. 

\subsection{Density Profiles, Enclosed Mass and Circular Velocity Profiles}

\begin{figure*}
	\includegraphics[scale=0.8]{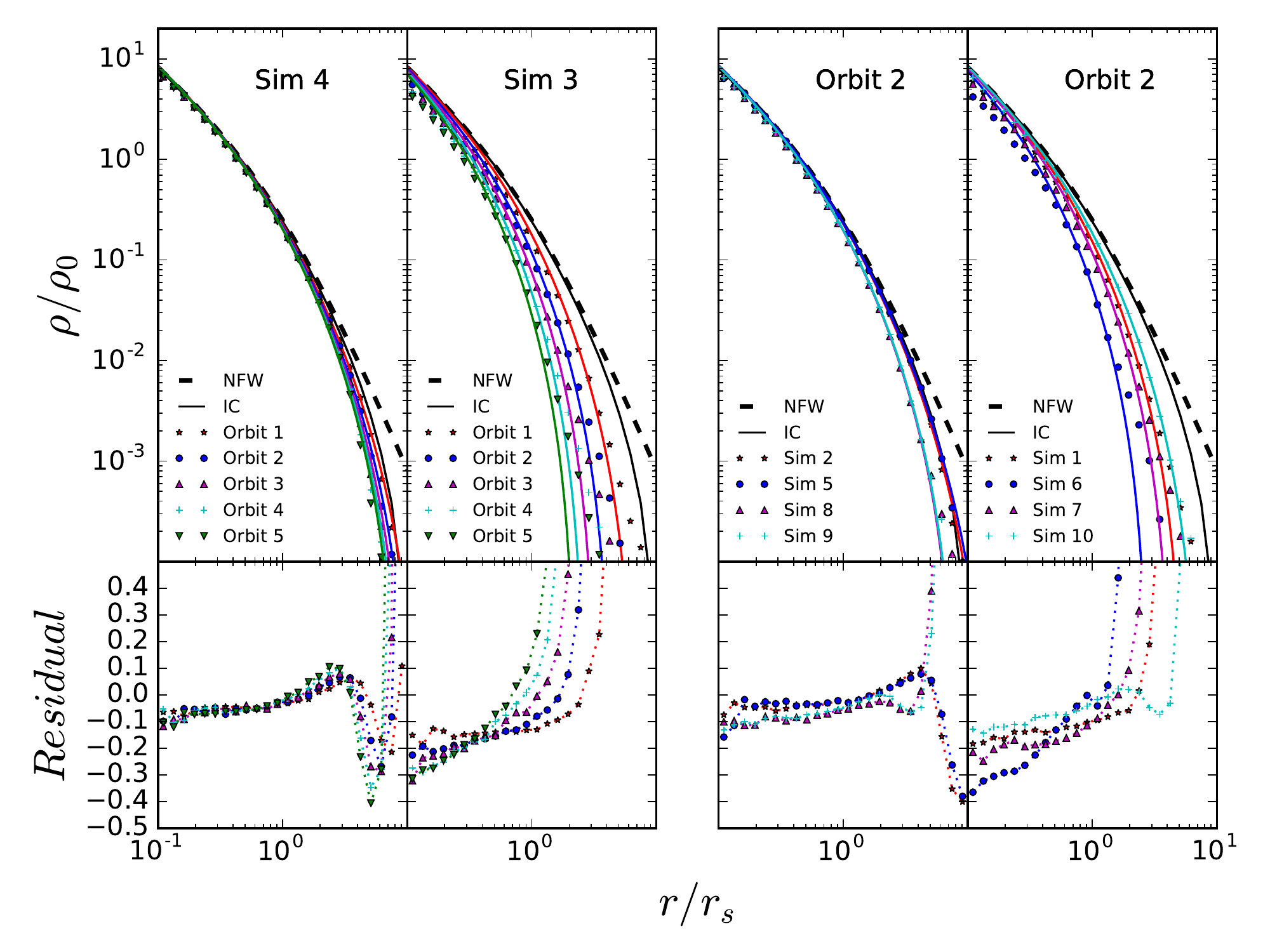}
	\caption{{\color{black}(Top panels) Density profiles of the bound satellite remnant. Simulation results are shown with points, and the best-fit energy-truncated model is shown with lines. The thick dashed curve shows an untruncated NFW profile, while the solid black curve shows the initial conditions at the start of the simulation. The two left panels show the first four orbits of Simulation 4 and Simulation 3. The two panels on the right show the profiles after two orbits for the simulations indicated. (Bottom panels) Relative residuals in density, ($\rho_{sim} - \rho_{model})/\rho_{model}$.}}
	\label{fig:sim_density}
\end{figure*}

The top panels of Fig.~\ref{fig:sim_density} show a comparison of the satellite density profiles from simulations (points) with the energy-truncated model (solid curves), where $Z_t$ has been fixed as described above. The thick dashed curve shows an untruncated NFW profile. 
{\color{black} We demonstrate how the profile changes in time for Simulations 4 and 3 in the two left-most panels, as well as how the different simulations compare in the two right-most panels. Note that Simulation 4 and Simulations 2, 5, 8 and 9 are the orbits with the slowest mass loss, and therefore the ones we might expect to be the most successfully described by our model.}
{\color{black}The bottom panels of Fig.~\ref{fig:sim_density} show the relative residuals in density, ($\rho_{sim}/\rho_{model} - 1$)}.
These are generally less than 10\%\ {\color{black} for the orbits with slower mass loss, or less than 20\%\ for orbits with faster mass loss. They are largest for Simulation 6, which loses mass extremely rapidly (see Fig.~\ref{fig:time_plots}), or at late times for Simulation 3. The residuals are generally} slightly larger at large radii, where an excess relative to the model is also visible in the top panels. Previous authors \citep[e.g.][{\color{black} and earlier references therein}]{penarrubia2008b, penarrubia2009,penarrubia2010} have noted a {\color{black} distinct} population of particles in the outer parts of tidally stripped systems, at radii too large to have crossed the satellite in the time since last pericentric passage. This may account for some of the excess seen here.

\begin{figure*}
	\includegraphics[scale=0.8]{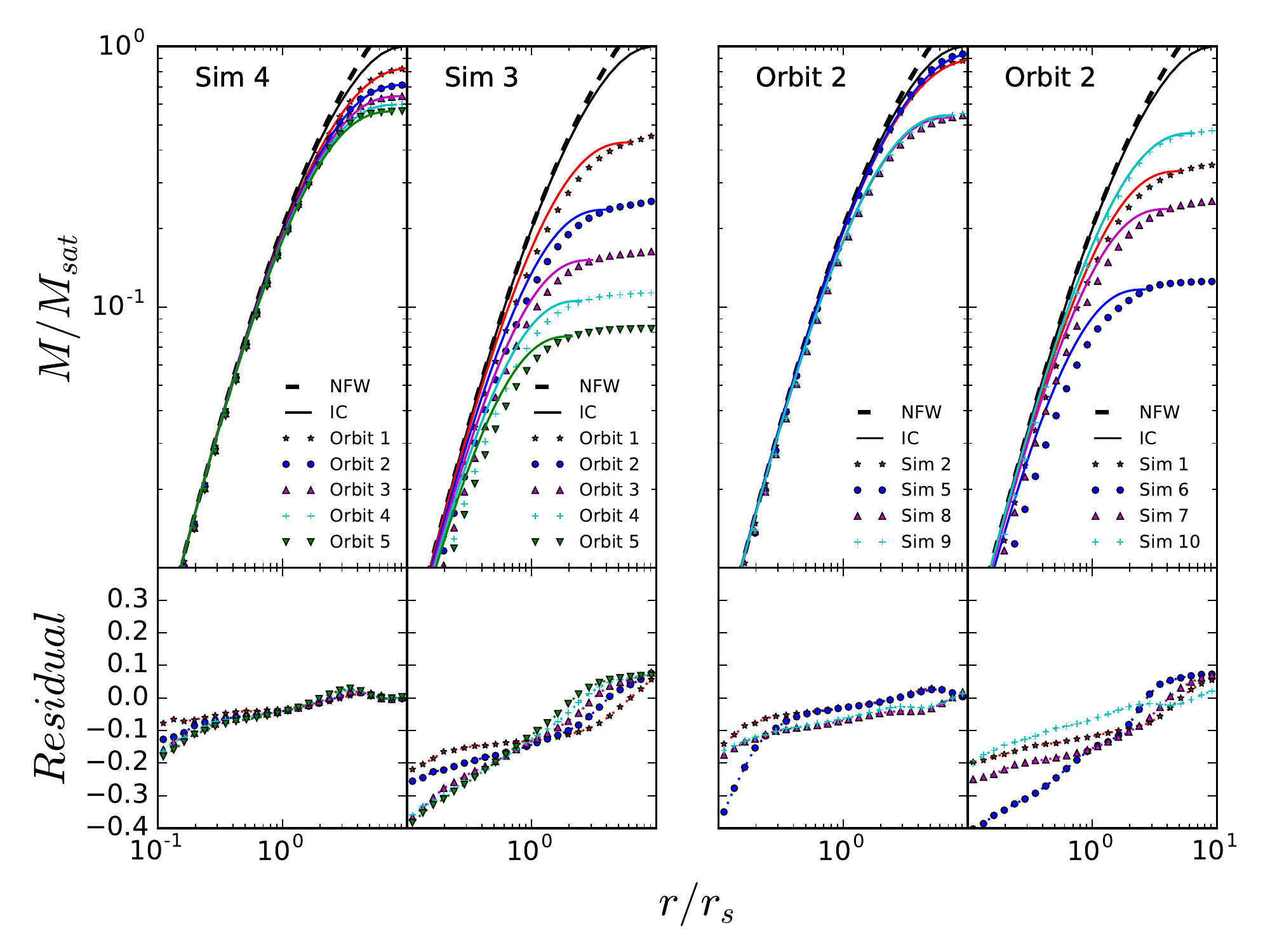}
	\caption{{\color{black}(Top panels) Cumulative mass profiles of the bound satellite remnant. Simulation results are shown with points, and the best-fit energy-truncated model is shown with lines. The thick dashed curve shows an untruncated NFW profile, while the solid black curve shows the initial conditions at the start of the simulation. The two left panels show the first four orbits of Simulation 4 and Simulation 3. The two panels on the right show the profiles after two orbits for the simulations indicated. (Bottom panels) Relative residuals in mass, ($M_{sim} - M_{model})/M_{model}$.  }}
	\label{fig:sim_mass}
\end{figure*}

\begin{figure*}
	\subfloat{\includegraphics[scale=0.8]{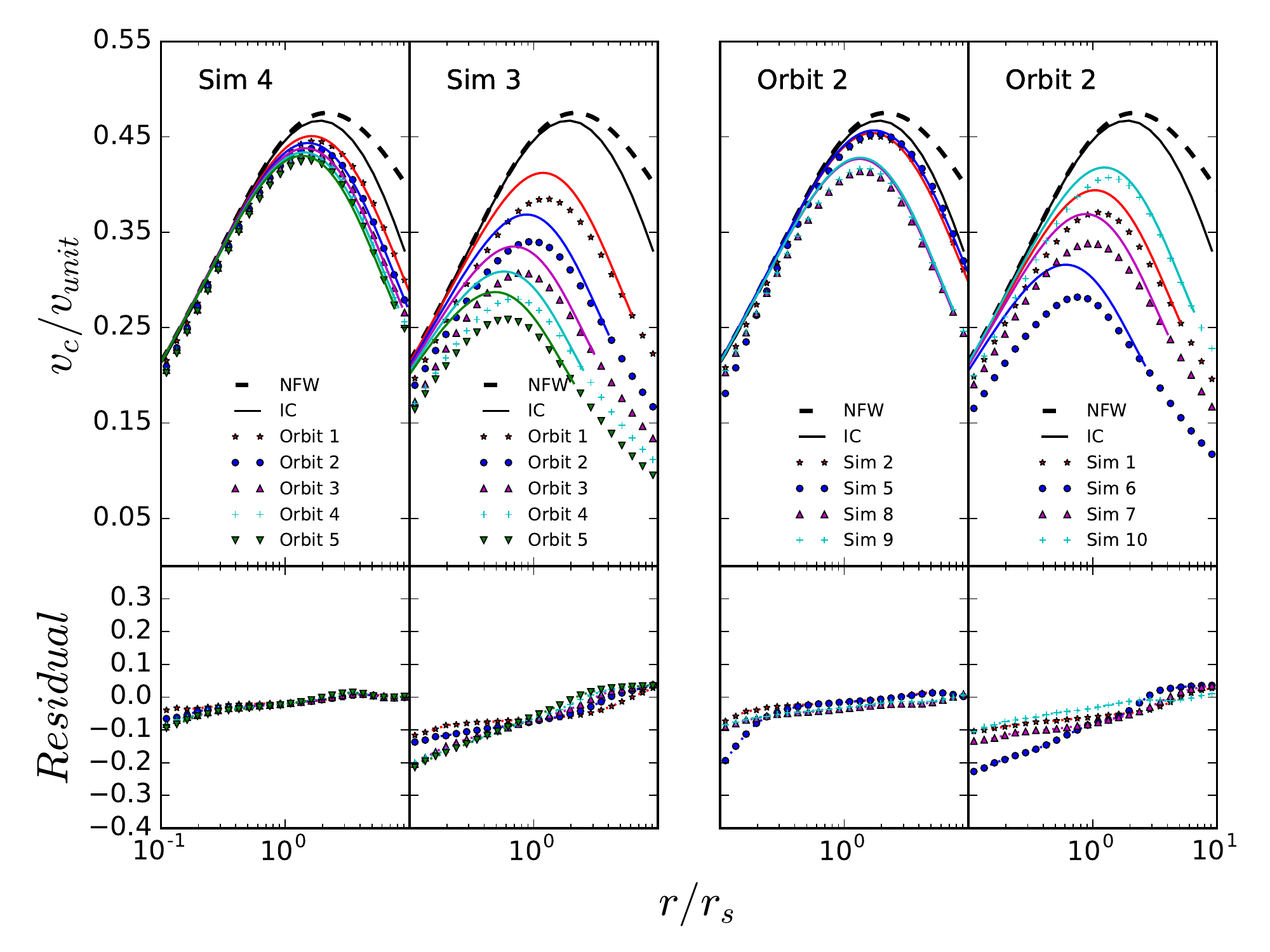}} \par
	\caption{{\color{black}(Top panel) Circular velocity of the bound satellite remnant as a function of radius. Simulation results are shown with points, and the best-fit energy-truncated model is shown with lines. The thick dashed curve shows an untruncated NFW profile, while the solid black curve shows the initial conditions at the start of the simulation. The two left panels show the first four orbits of Simulation 4 and Simulation 3. The two panels on the right show the profiles after two orbits for the simulations indicated. (Bottom panels) Relative residuals ($v_{sim} - v_{model})/v_{model}$.}}
	\label{fig:sim_velocity}
\end{figure*}

Fig.~\ref{fig:sim_mass} and Fig.~\ref{fig:sim_velocity} show a comparison of the enclosed mass and circular velocity profiles, respectively. Linestyles are as in Fig.~\ref{fig:sim_density}. Here too, residuals are generally at the 10--20\%\ level or less.
{\color{black}The largest deviations occur in Simulation 6, which loses mass the fastest. It should also be noted that Simulation 6 corresponds to an extreme, uncosmological orbit, with unusually large energy and extremely rapid mass loss, so it is less representative of realistic halo mergers.}
We {\color{black} also} note that our mass profiles and circular velocity profiles are constrained by the condition that the stripped mass fraction in the simulations matches the corresponding fraction in the models; a slightly different choice of $Z_t$ would have improved the agreement at some radii (e.g.~near the peak of the circular velocity curve in Simulation 1), at the expense of a slightly worse fit close to $r_{cut}$.

{\color{black} In general, the radius at which evaporation becomes important is not visible on the scales of Figs.~\ref{fig:sim_density} -- \ref{fig:sim_velocity} . The exception is Simulation 5, which has a very long orbital time scale. Thus, the deficit of mass at small radii for this simulation can be attributed to collisional effects.}

\subsection{Distribution Function and Moments of the Phase-Space Distribution}

Finally, we can attempt to compare the full DF of our simulated systems to the analytic energy-truncated model. We note {\color{black} that} the construction of the DF requires a frame in which to define velocities, and a set of particles to integrate over when calculating the potential. Thus the `distribution function' of a subsystem within a larger halo is a slightly problematic concept, relative to the usual definition for an isolated system. Here, to be concrete, we define velocities in the mean center-of-mass frame of the self-bound remnant, and calculate potential energies summing only over those particles that are bound.

\begin{figure*}
	\includegraphics[scale=0.8]{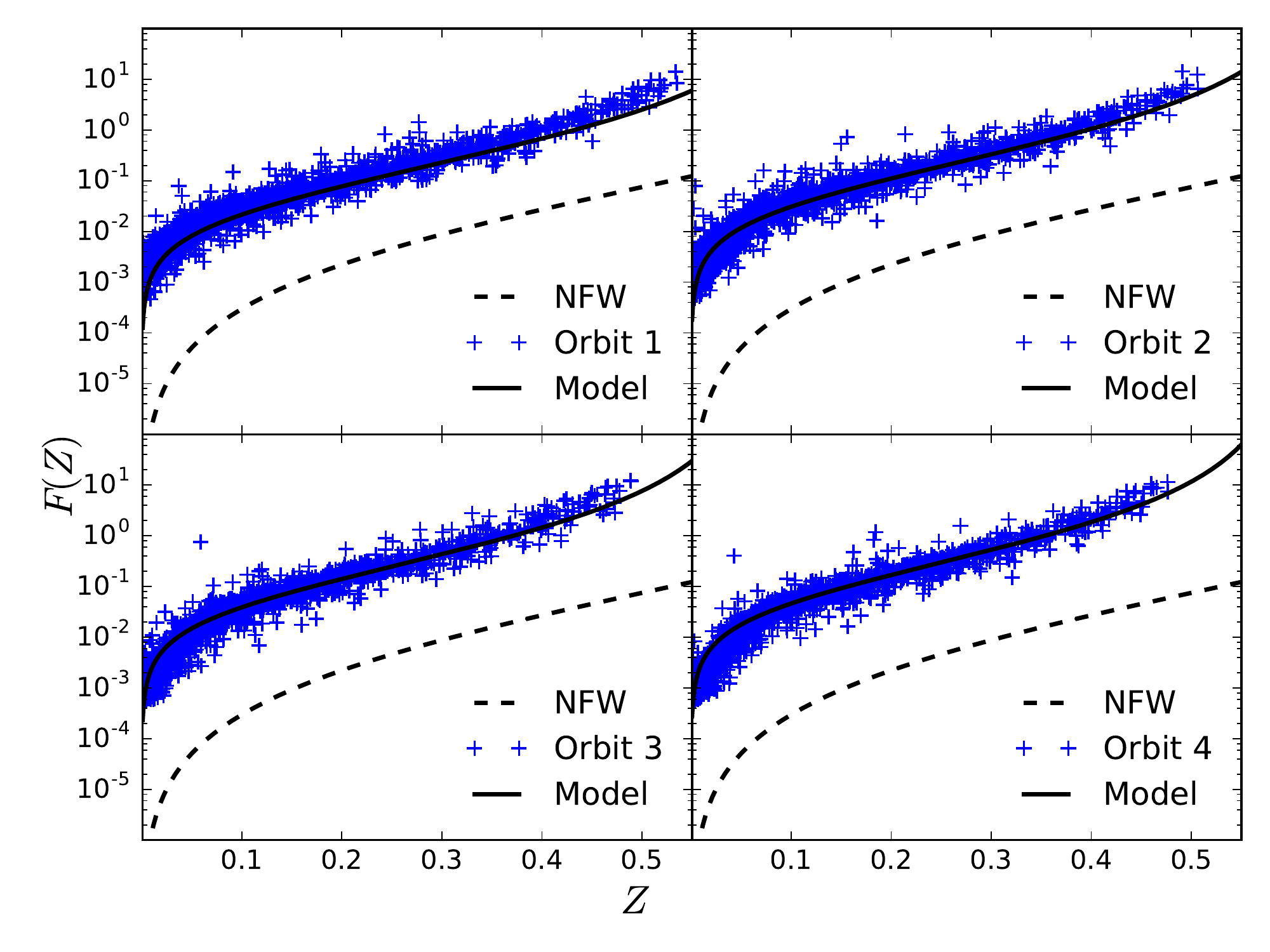}
	\caption{{\color{black}The DF as a function of the dimensionless relative energy $Z$, for the first four orbits of Simulation 4. The simulation results, binned in phase-space, are shown as crosses, while the predictions of the energy-truncated model are shown as solid curves. The dashed curves show the DF for an infinitely extended NFW profile.}} 
	\label{fig:sim_DF}
\end{figure*}

Given this convention, the DFs 
{\color{black}for the first four orbits of Simulation 4}
are shown in Fig.~\ref{fig:sim_DF}. Each point represents the number of particles in a bin of normalized {\color{black} relative } energy $Z$. This was found by binning the particles in $500$ equally sized bins in both radius and velocity. The phase-space density of each bin was then calculated as the number of particles divided by $16 \pi^2 r^2 v^2 dr dv$. {\color{black}  As usual, the relative energy is $\mathcal{E} = \Psi(r)  - v^2/2$, where $ \Psi(r )= -\Phi(r) + \Phi(r_{max})$, and $r_{max}$ is the radius of the remnant. Only bins with phase-space volume greater than $10^{-5}$ were plotted, to avoid numerical errors resulting from dividing by small numbers.}

The solid lines show the prediction of the analytic model from Section~\ref{sec:model} for comparison, while the dashed line shows the original (untruncated) NFW DF. Generally speaking, there is good agreement between the tidally stripped DFs and the analytic models for low {\color{black}relative energies, though it is difficult to compare the simulation to the model for large values of $Z$ (corresponding to particles at small radii and/or low energies), since the phase-space volume becomes very small in this limit.}

%Figure 9

\begin{figure*}
	\includegraphics[scale=0.6]{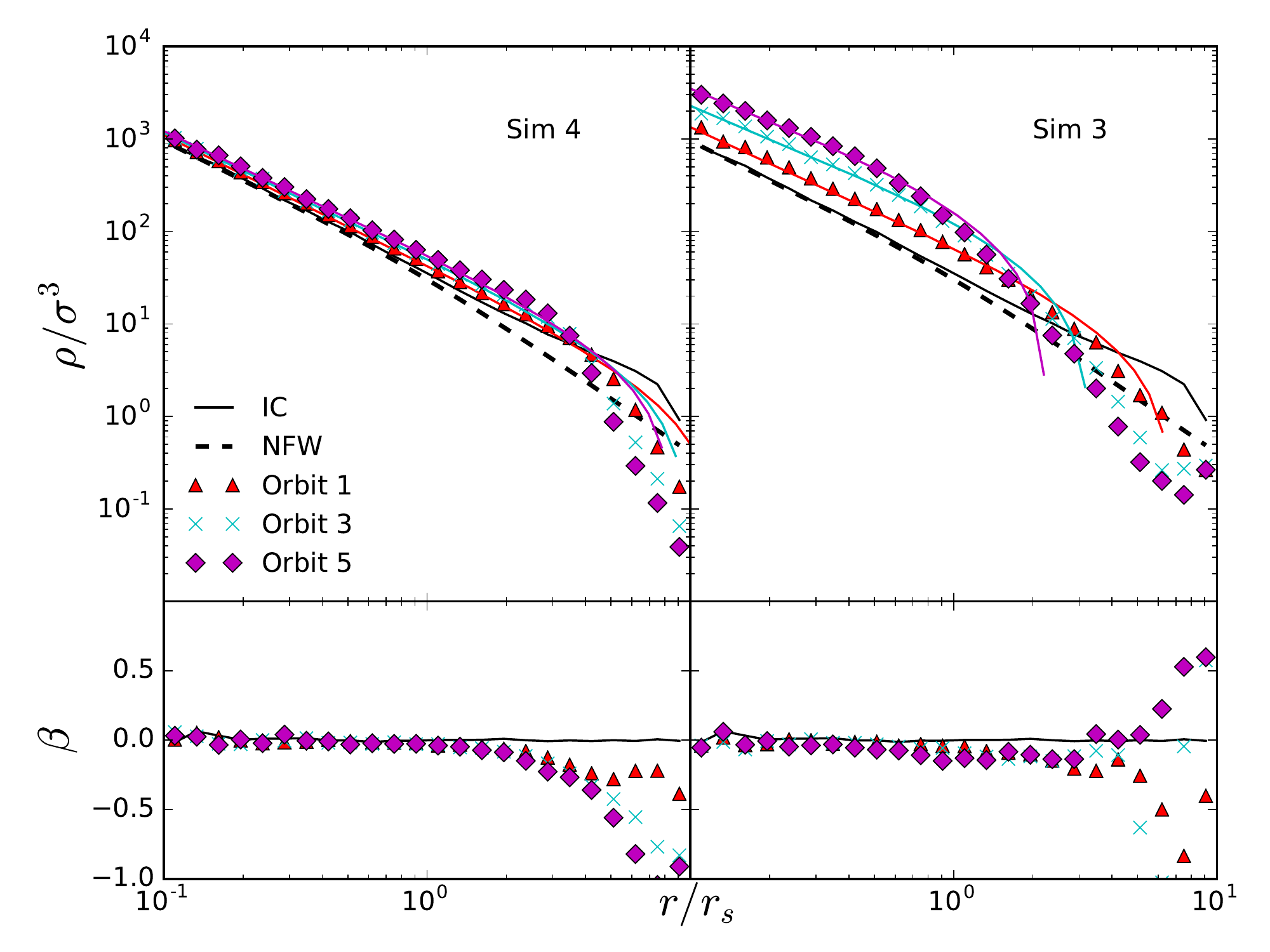}
	\caption{{\color{black}The pseudo-phase-space density (top) and anisotropy parameter (bottom) as a function of radius for Simulations 4 (left) and 3 (right). Density is in units $\rho_0$, and the velocity in units $v_{unit}=\sqrt{GM_{sat}/r_s}$. The thick dashed curve shows an untruncated NFW profile, while the solid black curve shows the initial conditions at the start of the simulation. Simulation results are shown with points, and the analytic model with lines.}} 
	\label{fig:pseudo}
\end{figure*}

{\color{black}
In Fig. \ref{fig:pseudo} we also show the evolution of the pseudo-phase-space density for Simulations 3 and 4 (top panel). As first noted by \cite{taylor2001b}, the spherically averaged pseudo-phase-space density, $\rho/\sigma^3$, of isotropic, cosmological halos appears to follow a simple power law as a function of radius. The dashed line shows this power-law for an infinitely extended NFW profile. For the truncated analytic models (solid lines), we can calculate the velocity dispersion from the probability distribution function for velocities, $P(v) \propto f(\mathcal{E})v^2/\rho(r)$.  The analytic models also follow a power law out to around the truncation radius, but with a flatter slope. The relative increase in pseudo-phase-space density at large radii is expected, since energy truncation reduces the number of particles with large velocities, and thus the velocity dispersion, in these regions. A similar flattening of the slope has also been seen in tidally truncated subhalos from self-consistent cosmological simulations \citep{veraciro2014}. 

Interestingly, while the simulation results (points) match the analytic models extremely well at small radii, they deviate from them systematically at large radii. The bottom panel shows the likely reason for this discrepancy:  the tidally stripped simulations are not isotropic in their outer regions. Plotting the anisotropy parameter $\beta = 1 - (\sigma_\theta^2 + \sigma_\phi^2)/2 \sigma_r^2$ as a function of radius, we see that the analytic model and simulation results in the top panel begin to differ at radii where $\beta$ is becoming significantly different from zero, suggesting that our analytic model fails to match the outer parts of the simulations primarily because the assumption of isotropy is no longer valid there.}

\begin{figure*}
	\includegraphics[scale=0.7]{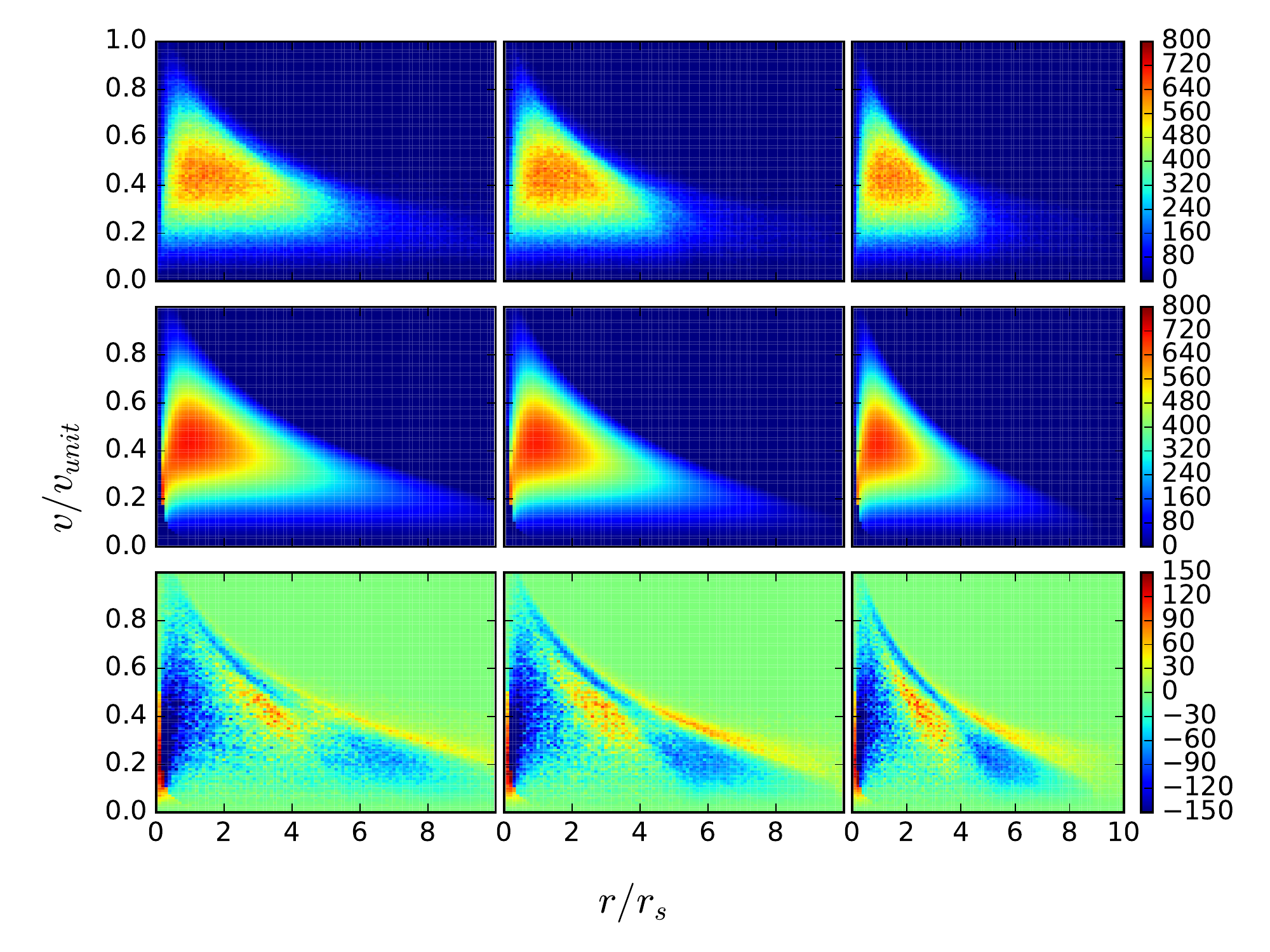}
	\caption{Number of particles {\color{black} as a function of radius and total velocity}. The figure shows {\color{black}Simulation 4} results (top row), the analytic model (middle row), and the difference between the two (bottom row), at each of the first three apocentric passages (three columns).}
	\label{fig:sim_phase}
\end{figure*}

To get a better sense of where the simulations differ from the models in {\color{black} the full} phase space, we can also bin the DF in radius and (total) velocity. Fig.~\ref{fig:sim_phase} shows this {\color{black} 2-D} distribution for the first three orbits of Simulation 4 (top three panels), the corresponding analytic models (middle panels), and the differences between the two (lower panels). Generally speaking, there is good agreement between the models and the simulations {\color{black} in this projection of the full phase space}. The tidally stripped satellites show a slight excess of particles at large radii and large velocities, and this excess may even have a caustic-like structure to it. It seems likely that this is related to the excess component at large radii {\color{black} noted by earlier authors and discussed in the previous section}. \citet{penarrubia2009} found that this outer material, while still bound, is on its way out of the system and is mostly lost on the next orbit. Such non-equilibrium effects would not be captured by our analytic truncation model. There is also a slight deficit of particles at small radii and low velocities. Here, relaxation may play a role in scattering particle energies, and depleting the lowest-energy orbits.

%%%%%%%%%%%%%%%%%%%%%%%%%%%%%%%%%%%%%%%%%%%%%%%%%%
%%%%%%%%%%%%%%%%%%%%%%%%%%%%%%%%%%%%%%%%%%%%%%%%%%
%%%%%%%%%%%%%%%%%%%%%%%%%%%%%%%%%%%%%%%%%%%%%%%%%%
	
\section{Discussion and Conclusions} \label{sec:conc}

Dark matter halos play a central role in the current picture of cosmological structure formation, and their properties and evolution determine many of the broad trends in the properties of galaxies, galaxy groups and galaxy clusters. Our understanding of halo formation and evolution is still limited, however; many basic questions, such as the origin of the universal density profile or the concentration-mass relation, have not yet been answered fully. Progress in understanding halo properties will come from a combination of fully realistic, large-volume cosmological simulations, together with  simpler, idealized simulations of individual halos. Our original goal, in this work, was to develop a better algorithm for generating initial conditions for the latter.

We have found through experimentation that two different approaches yield similar results: either truncating an NFW profile abruptly at some radius $r_{cut}$, and then iteratively removing unbound particles until convergence is reached, or using an analytic NFW distribution function truncated at some energy and shifted such that the resulting distribution function is continuous at $\mathcal{E} = 0$. The latter technique, inspired by the King model \citep{king1966} {\color{black} and proposed previously by \cite{widrow2005}}, produces almost the same density profile as the former, and allows us to construct simple analytic models for spatially finite systems whose central regions resemble NFW profiles. For generating initial conditions in practice, we have used the first method; our tests show that these models are extremely stable, and thus well suited to the study of isolated systems.

For either of these solutions, the density profile drops off steeply, reaching zero at a finite truncation radius. This behaviour is familiar from numerical studies of tidally stripped halos, starting with \citep{hayashi2003}.  Pursuing the suggestion of a connection between the two, we have simulated the tidal evolution of satellite subhalos in the potential of a larger system. We find that the resulting tidally stripped remnants match our theoretical models at the 10--20\%\ level or better in density profile, enclosed mass and/or circular velocity profile. In phase space they also look similar, particularly when we plot phase-space density {\color{black} projected} in terms of radius and {\color{black} total} velocity. Some of the minor differences may relate to the presence of unrelaxed material close to the outer edge of the system, as discussed by \citet{penarrubia2009}, or to relaxation effects near its centre. The origin and full significance of these differences will require further work to clarify. Nonetheless, overall our analytic model provides a good first approximation to the detailed distribution function of tidally stripped systems.

This result is actually slightly surprising. Tidal mass loss is a complex process, and has been investigated by many authors previously, using different approaches and approximations. \citet{taylor2001}, for instance, developed a 1-D model for tidal stripping, calculating the average effects of tidal heating and expansion, as well as adiabatic cooling, within spherical shells, and removing the mass outside an instantaneous tidal limit over a timescale equal to the instantaneous (angular) orbital period, $t_{orb} = 2\pi/\omega$. A limitation of this model, in the present context, is that it did not specify the actual density profile of the satellite at any one time, but only the net effect of heating and cooling on mass loss. \citet{Benson2002} considered a similar model, but with several corrections, notably using the shorter of the angular orbital period and the radial infall time $R/v_R$ as the timescale for mass loss. They found this gives better agreement with mass loss rates from simulations for very radial orbits. Analytic models of a type similar have also been developed by several other groups \cite[e.g.][]{zentner2003,oguri2004,vandenbosch2005,zentner2005}.  

\citet{Kampakoglou2007} considered the problem of mass loss in more detail, calculating the effects of heating {\color{black} on circular orbits} more precisely using particles to sample the distribution function in energy and angular momentum, as well as for particle orbits of different inclination. They found that by successively heating and removing unbound particles, they predicted a profile similar to the one seen in their simulations, at least in the limit of weak tidal fields. This may provide the first hint that modelling mass loss in energy space can naturally explain the profile of tidally stripped cosmological halos. 

Given that they were trying to establish simpler, semi-analytic prescriptions for tidal mass loss, none of these approaches considered the full complexity of the problem. In reality, the tidal boundary is non-spherical even for a spherical system on a circular orbit; for a general orbit it is also time-varying, and the timescales for relaxation and mass removal in the outer parts of the satellite are unfortunately very close to the orbital timescales. Furthermore, real halos are usually triaxial, and may have their spin partly coupled to their orbital angular momentum. Thus a general, predictive and highly accurate model for tidal mass loss seems a distant prospect.

Nonetheless, some simple patterns do emerge. \citet{hayashi2003} first pointed out that all tidally truncated NFW profiles look the same, and can be fit by a single additional parameter; they showed that mass loss in their simulations is essentially a 1-dimensional sequence, a result later confirmed for a wider range of profiles by \citet{penarrubia2010}. Our results now connect this stripping sequence to a simple cutoff and shift in the underlying distribution function, parameterized by an increasing truncation energy $\mathcal{E}_t$. 
Exactly how heating, relaxation, and unbinding conspire to produce this simple result is unclear {\color{black}. A hint comes from the work of \cite{choi2009}, who show that individual particles in tidally stripped systems do or don't become unbound based primarily on their energy, rather than their angular momentum. Thus, a simple cut in energy may provide a fairly accurate description of the mass-loss process. 

Producing a more detailed description of tidal mass loss} will be the focus of future work. In the interim, the models presented here provide a new, physically motivated way of  generating initial conditions for isolated halos or tidally truncated subhalos. {\color{black} Applications include studies of tidal stream formation \citep[e.g.][and references therein]{amorisco2015}, disk heating \citep[e.g.][and references therein]{moetazedian2016}, mergers \citep[e.g.][and references therein]{carucci2014},  and dwarf galaxy evolution \citep[e.g.][and references therein]{tomozeiu2016}.} A python code for generating ICs {{\color{black} using our iterative method is available on-line}. Its use is described in the appendix.  

{\color{black} Finally, we note several limitations to this work. We have studied the tidal evolution of idealized, spherical, isotropic halos, with an NFW density profile and corresponding DF. Halos in self-consistent cosmological simulations differ from this idealized case in a number of ways. It is tempting to look to self-consistent simulations to try to understand tidal mass loss in a more realistic situation; \cite{springel2008}, for instance, show the density profiles of tidally truncated subhalos resolved in the Aquarius simulations with $10^5-10^6$ particles or more. Broadly speaking they resemble our models, with a profile similar to field halos (i.e. close to an NFW or Einasto profile) in the inner parts, and truncated abruptly in the outer parts. Unfortunately, in self-consistent simulations the exact profile and distribution function of a subhalo will depend on the group-finder used. Group-finders use various different criteria for defining the boundary of a subhalo, producing slightly different results (e.g.~\citealt{muldrew2011}; see \citealt{onions2012} for a general review). Establishing precisely which particles are or aren't associated with a subhalo, in a region dominated by the background potential of the main system, is essential impossible. Thus, the idealized simulations presented here actually provide more reliable information about the behaviour of stripped systems close to the tidal boundary, and may even help to evaluate different group-finding schemes in realistic simulations. 

The isotropic NFW models considered in this work do not necessarily provide the most accurate description of cosmological halos, either. The highest-resolution simulations indicate that an Einasto profile is a better fit to the density profile \citep[e.g.][]{navarro2010, klypin2016}, but real halos are also anisotropic, triaxial, and have more complicated correlations between shape and anisotropy \citep[e.g.][]{wojtak2013}. There are also several proposed models for the `true' DF of dark matter halos, based on maximizing entropy under various constraints, or other arguments (e.g.~\citealt{Hjorth2010,Pontzen2013,Beraldo2014}; see \citealt{halle2017} for a discussion). Our focus here is not on which of these theoretical models is correct, but on the generic effects of tidal mass loss on any DF. Our analysis is general and may be applied to these theoretical models, or to any other model with an explicit DF.

Many real systems of interest also have a separate baryonic component that is important, if only as a tracer of dynamics and mass loss. Various authors have considered the density profiles and/or phase-space distributions of luminous stars within a surrounding, tidally limited dark matter potential, and their evolution through tidal interactions \citep[e.g.][]{mashchenko2004,mashchenko2005a,mashchenko2005b,penarrubia2008b,sales2010,kazantzidis2011,amorisco2011}. Our focus here is simply on understanding the phase-space evolution of the dark matter particles themselves. We leave further discussion of the evolution of self-consistent, two-component systems to future work.}

%%%%%%%%%%%%%%%%%%%%%%%%%%%%%%%%%%%%%%%%%%%%%%%%%%
%%%%%%%%%%%%%%%%%%%%%%%%%%%%%%%%%%%%%%%%%%%%%%%%%%
%%%%%%%%%%%%%%%%%%%%%%%%%%%%%%%%%%%%%%%%%%%%%%%%%%
	
	\section*{Acknowledgements}
NED acknowledges support from the government of Ontario, through an OGS award. JET acknowledges financial support from NSERC Canada, through a Discovery Grant. The authors also wish to thank Frank van den Bosch and the anonymous referee for useful comments.

	\appendix
{\color{black}
	\section{ICICLE -- A Code for Generating Isolated Initial Conditions}} \label{sec:Appendix}

The authors have created a {\color{black}publicly available Python package, \textsc{Icicle} (Initial Conditions for Isolated CoLlisionless systEms),} which can create stable initial conditions for spherical, isotropic, collisionless systems with various density profiles. The code currently supports NFW (either exponentially truncated, abruptly truncated, or truncated {\color{black} using the iterative method} described in Section~\ref{sec:ICs}), King, Hernquist and Einasto profiles.

\subsection{Files}

The main part of the program is in the code ``ICICLE.py". The content of this code is described thoroughly in Section \ref{sec:overview}; {\color{black} briefly,} given a model it returns positions and velocities of $n$ particles within that distribution.

For each model there is an additional file named ``ICs\_\emph{model}.py". These files contain information about the specific model (the density profile, cumulative mass distribution, gravitational potential and distribution function). {\color{black}The user can specify parameters by use of a parameter file.}

\subsection{Output}

{\color{black}
	The output is written to the filename specified. The first line of the text file has the following information: the number of particles, the mass of each particle and the value of the gravitational constant. Subsequent lines display the particle number (indexed from 0), the x, y and z positions, and the x, y, and z velocity components.
}

%%%%%%%%%%%%%%%%%%%%%%%%%%%%%%%%%%%%%%%%%%%%%%%%%%%%%%%%

\subsection{Positions and Velocities} \label{sec:overview}

In this section we outline the steps needed to select positions and velocities given an isotropic density profile, in a manner similar to \cite{kazantzidis2004}. First, the radial distance {\color{black} for each particle} is selected using the cumulative mass distribution, and then {\color{black} a direction for the position vector} is selected assuming spherical symmetry. Next, an energy is selected from the energy distribution. Once the position and energy of the particle have been determined, it is straightforward to calculate the velocity of the particle. Finally, the velocity direction is chosen isotropically.

%%%%%%%%%%%%%%%%%%%%%%%%%%%%%%%%%%%%%%%%%%%%%%%%%%%%%%%%
\subsubsection{Positions}

{\color{black} The radius for each particle is chosen in such a way as to reproduce} the mass distribution. Consider a density profile where the mass within a given radius $r$ is $M(<r)$, and the largest radius is $r_{max}$. The mass fraction interior to $r$ is then:
\begin{equation}
F_M(<r) = \dfrac{M(<r)}{M(r_{max})} \,\,\, .
\end{equation}
{\color{black} If we choose a random variate $x$ uniformly on the interval $[0,1]$, we can then set $F_M(<r) = x$ and solve for $r$ to find a corresponding radius. In the limit of a large number of particles, the resulting set of radii will reproduce the desired density profile. Given a radius, we can then choose a direction isotropically, by choosing a random point on the surface of a unit sphere.}

%%%%%%%%%%%%%%%%%%%%%%%%%%%%%%%%%%%%%%%%%%%%%%%%%%%%%%%%
\subsubsection{Calculating The Distribution Function}

The distribution function, $f(\mathcal{E})$, is given by Eddington's formula:
\begin{equation} \label{eq:eddington}
f(\mathcal{E})=\dfrac{1}{\sqrt{8}\pi^2}\left[ \int_0^\mathcal{E} \dfrac{1}{\sqrt{\mathcal{E}- \Psi}}\dfrac{d^2 \rho}{d \Psi^2} d\Psi +\dfrac{1}{\sqrt{\mathcal{E}}}\left( \dfrac{d\rho}{d\Psi}\right)_{\Psi=0}  \right] \,\,\, .
\end{equation}
Here $\mathcal{E}$, $\Psi$ and $\rho$ are the relative energy, the relative potential and the density, respectively. {\color{black} The second, boundary term is equal to $f(0)$ in general (since the first term evaluates to zero when $\mathcal{E} = 0$), so it vanishes for any system with $f(0)=0$.} 

Consider the term $d^2 \rho/d \Psi ^2 $. This can be expressed as:
\begin{equation} \label{eq:drhodPsi}
\begin{gathered}
\dfrac{d^2 \rho}{d \Psi^2} = \left( \dfrac{d\Psi}{dr}\right)^{-2} \left[ \dfrac{d^2 \rho}{dr^2} -  \left( \dfrac{d\Psi}{dr}\right)^{-1}   \dfrac{d^2\Psi}{dr^2} \dfrac{d\rho}{dr}\right] \\
{\rm or,\ since\ \ \ \ \ \ }\dfrac{d\Psi}{dr} = - \dfrac{GM}{r^2} \\
{\rm and\ \ \ \ \ } \dfrac{d^2\Psi}{dr^2} = \dfrac{2GM}{r^3} - 4\pi G \rho\\ 
\dfrac{d^2 \rho}{d \Psi^2} = \left( \dfrac{r^4}{G^2M^2}\right)\left[ \dfrac{d^2 \rho}{dr^2} + \left( \dfrac{r^2}{GM}\right) \left[  \dfrac{2GM}{r^3} - 4\pi G \rho\right]\dfrac{d\rho}{dr}\right]\,\,\, .
\end{gathered}
\end{equation}
{\color{black} This form is more convenient in cases where $\rho$ is an analytic function of radius, as it avoids having to take potentially noisy numerical derivatives.}

\subsubsection{Choosing From the Distribution Function}

The probability that a particle located at radius $r$ has a relative energy $\mathcal{E}$ is proportional to $f(\mathcal{E}) \sqrt{\Psi - \mathcal{E}} $, {\color{black} with a maximum energy of $\mathcal{E}_{max}=\Psi(r)$. }

Thus, the cumulative distribution function (CDF) $F(<\mathcal{E})$, for a particle at position $r$ with relative potential energy $\Psi(r)$, is:
\begin{equation}
F(<\mathcal{E})=\dfrac{\int_0^{\mathcal{E}} f(\mathcal{E}) \sqrt{\Psi - \mathcal{E}}  d \mathcal{E}}{\int_0^{\Psi} f(\mathcal{E}) \sqrt{\Psi - \mathcal{E}}  d \mathcal{E}} \,\,\, .
\end{equation}
{\color{black} Choosing a random variate $y$ uniformly on the interval $[0,1]$, we can set $F(<\mathcal{E})) = y$ and solve for $\mathcal{E}$ to find the corresponding relative energy. In practice, we do this by linear interpolation of a numerically determined CDF.}

\subsubsection{Velocities}

{\color{black} Once a position and a relative energy have been chosen for the particle, the velocity} magnitude can be calculated from:
\begin{equation}
\mathcal{E}=\Psi-\dfrac{1}{2}v^2 \,\,\, .
\end{equation}
{\color{black} Finally, a direction for the velocity vector can be chosen isotropically in the same way as described for the positions.} 

\subsection{Profile options and documentation}

{\color{black} The code, a list of supported profile types, and full documentation are available as a package, \textsc{Icicle}, at the URL:  https://github.com/ndrakos/ICICLE}.

%%%%%%%%%%%%%%%%%%%%%%%%%%%%%%%%%
%%%%%%%%%%%%%%%%%%%%%%%%%%%%%%%%%%
%%%%%%%%%%%%%%%%%%%%%%%%%%%%%%%%%%

	\bibliographystyle{mnras}
	\bibliography{TidalStripBib}
	
	% Don't change these lines
	\bsp	% typesetting comment
	\label{lastpage}
\end{document}